\documentclass[preprint,prd,showpacs,superscriptaddress]{revtex4}
\usepackage{amssymb,amsmath,graphicx,amscd}

\begin{document}

\title{Quantum Stress Tensor Fluctuations and Primordial Gravity Waves}

\author{Jen-Tsung Hsiang}\email{cosmology@gmail.com}
\affiliation{Center for Particle Physics and Field Theory, Department of Physics, 
Fudan University, Shanghai 200433, China}
\author{L. H. Ford}\email{ford@cosmos.phy.tufts.edu}
\affiliation{Institute of Cosmology, Department of Physics and
Astronomy,
 Tufts University, Medford, MA 02155, USA}
\author{Kin-Wang Ng}
\email{nkw@phys.sinica.edu.tw}
\affiliation{Institute of Physics,
Academia Sinica, Nankang, Taipei 11529, Taiwan}
\author{Chun-Hsien Wu}
\email{chunwu@scu.edu.tw}
\affiliation{Department of Physics, Soochow University, \\
70 Linhsi Road, Shihlin, Taipei 111, Taiwan}

\begin{abstract}
We examine the effect of the stress tensor of a quantum matter field, such as the electromagnetic field, on
the spectrum of primordial gravity waves expected in inflationary cosmology. We find that the net effect is
a small reduction in the power spectrum, especially at higher frequencies, but which has a different form
from that described by the usual spectral index. Thus this effect has a characteristic signature, and is in
principle observable. The net effect is a sum of two contributions, one of which is due to quantum fluctuations
of the matter field stress tensor. The other is a quantum correction to the graviton field due to coupling to
the expectation value of this stress tensor. Both contributions are sensitive to initial conditions in the very
early universe, so this effect has the potential to act as a probe of these initial conditions.
\end{abstract}


\maketitle

\baselineskip=16pt

\section{Introduction}
\label{sec:intro}

The possible roles of quantum stress tensor fluctuations of a conformal matter field in inflationary cosmology were explored in
Refs.~\cite{WKF07,FMNWW10,WHFN11}. The basic hypothesis is that vacuum fluctuations of the energy
density, or other stress tensor components, during inflation can lead to additional density or tensor 
perturbations beyond those due to the nearly Gaussian fluctuations of the inflaton or graviton fields, which are 
the dominant source of perturbations in inflationary models.  One of the usual
features of inflationary models is the relative lack of dependence upon initial conditions. This arises because adequate
inflationary expansion will redshift and dilute pre-existing matter, so a generic quantum state tends eventually to become
indistinguishable from the Bunch-Davies vacuum of de Sitter spacetime. However, the effects of quantum stress tensor 
fluctuations exhibit a strong dependence upon the initial conditions, and tend to grow with increasing duration of the
inflationary expansion. In particular, the contributions to the density perturbations~\cite{WKF07,FMNWW10} and the
tensor perturbations~\cite{WHFN11} are proportional to the scale factor change between an initial time and the end
of inflation. At first sight, this violates Weinberg's theorem~\cite{W-thm}, which states that loop corrections can grow at 
most logarithmically in the scale factor during inflation. However, a more careful analysis shows that there is in fact no
contradiction. One picks a perturbation with a given wavelength at the present time, and then traces it back to the
initial time, when it becomes a perturbations with very short wavelength, which is inversely proportional to the scale 
factor change during inflation.
However, the magnitude of stress tensor fluctuation effects increases as the relevant length scale decreases. 
The stress tensor fluctuation contribution can
typically be written as a double integral over conformal time of the  stress tensor correlation function.
Thus, increasing the
duration of inflation leads to increasing values of the integrals of the  stress tensor correlation function, because shorter
proper wavelength modes are giving the dominant contribution, not because anything is accumulating as inflation
progresses. Another way of saying this, is that the the dominant contribution is always large, but is not growing in time.
Thus, it is misleading to refer to these contributions as ``secular terms".

The main purpose of the present paper is to perform a more careful analysis of the tensor perturbations due
to stress tensor fluctuations which was begun in Ref.~\cite{WHFN11}. The analysis in this reference was criticized 
by  Fr{\"o}b, {\it et al.}~\cite{FRV12}, who noted correctly that an important contribution was omitted. In addition to the
passive fluctuations of the graviton field due to stress tensor fluctuations, there is also a correction to the active
fluctuations of the graviton field which is of the same order. Fr{\"o}b, {\it et al.} propose a method for calculating the 
combined effect of both contributions which leads to a result which does not depend upon the duration of inflation,
and infer that the two contributions have cancelled one another. In the present paper, we propose a rather
different approach in which each contribution is computed separately, with initial conditions  imposed by a switching
function.  This function describes the switch-on of the coupling of the matter stress tensor with the graviton field. 
We do not find cancellation between the passive and active fluctuation effects, but rather a nonzero one loop
correction to the graviton power spectrum.

The outline of the paper is as follows: Section~\ref{sec:switch} will discuss averaging of quantum
fields with switching functions of time. Section~\ref{S:free-gravitons} will review free gravitons in an expanding universe,
and Sec.~\ref{sec:matter} will introduce some formalism for describing the effects of a quantum matter field.
Section~\ref{S:P22} will compute the correction to the graviton power spectrum coming from matter stress
tensor fluctuations. These are the passive fluctuations of gravity driven by the fluctuating stress tensor. The part
of the power spectrum coming from loop corrections to the graviton field, the modified active fluctuations, is treated
in Sec.~\ref{S:P13}. The combination of both effects is discussed in Sec.~\ref{S:combined}, and shown in our
approach to be nonzero, and dependent upon the details of the switching function. In general, it has the effect of
slightly reducing the overall power spectrum. Some numerical estimates, and the possibility of observing this reduction
will be discussed. Our results will be summarized and discussed in Sec.~\ref{S:sum}.

\section{Switching Functions in Quantum Field Theory}
\label{sec:switch}

In interacting quantum field theory calculations of scattering amplitudes, it is common to appeal to adiabatic switching
of the coupling constants of the theory. This allows the theory to be non-interacting in the past and in the future, and
allows a well defined $S$-matrix to connect the asymptotic states. Here the term ``adiabatic" means that the scale
of any time dependence in the coupling constants is long compared to any other time scales in the theory. The details
of the time dependence usually need not be specified. 

There is a different type of switching required in treatments of the fluctuations of quantum stress tensor operator.
The fluctuations of a local field operator are not defined, as the moments of such an operator will diverge if they
are  not zero. However, the time average or spacetime average of a local stress tensor operator does have finite
moments, and can have a meaningful probability distribution in a given quantum state, as discussed in 
Refs.~\cite{FFR10,FFR12,FF15}. Here we briefly review the basic ideas, and describe the role which they will play
in the present paper. Let $T$ be a stress tensor operator component, such as energy density, which has been 
renormalized to have  finite expectation values. In Minkowski spacetime, this means a normal ordered operator. 
In curved spacetime, it means an operator which has been regulated and renormalized, as will be reviewed in 
Sec.~\ref{sec:matter}. Although the local operator $T$ may have finite expectation values in physically realizable
states, its fluctuations are not well defined because its higher moments, $\langle T^n \rangle$ for $n \geq 2$ typically
diverge. The physical implication of this divergence is that the fluctuations of a local operator, such as energy density
at a single spacetime point are infinite and hence not meaningful. However, if we average $T$ over a finite time or
spacetime interval, then the fluctuations do become well defined.

Let
\begin{equation}\label{E:aver}
\bar{T} = \int_{-\infty}^\infty s(t) \, T(t)\, dt 
\end{equation}
be the average of $T$ at a given spatial point with respect to a smooth ($C^\infty$) switching function $s(t)$, which satisfies
$s(t) \rightarrow 0$ for $t  \rightarrow \pm \infty$. Now the moments of $\bar{T}$ become finite, and a meaningful probability 
distribution may be defined~\cite{FFR10,FFR12,FF15}. The effect of the time averaging described by  $s(t)$, is to
suppress the contribution of high frequency modes to the moments. Averaging in space alone at a single time is not adequate
to produce this suppression, but averaging in both space and time has qualitatively the same effect as does time averaging alone.
One may view the averaging of a stress tensor operator as describing its measurement process in a given physical situation,
where the details of the situation define the switching function. An example of this was discussed in Ref.~\cite{HF16}, which
treats the effects of vacuum radiation pressure fluctuations on quantum particles near a potential barrier. Here the duration
of the switching is taken to be determined by the time which the particle spends in the vicinity of the barrier, and the form of the
switching function to be defined by the shapes of both the barrier and the wavepacket of the particle. Similar ideas were
applied in Ref.~\cite{HF15} to the fluctuations of the quantized electric field. Most of the above discussion of averaged stress
tensor operators, which are quadratic in the quantum field operator, applies to linear quantum fields, such as the electric field.
The primary difference is that the stress tensor probability distribution is sensitive to the functional form of the switching
function, whereas the probability distribution for the electric field is always a Gaussian, and depends primarily upon the
width of the switching interval. In Ref.~\cite{HF15}, it was shown that if this switching time is taken to be the time which a charged
particle spends in the vicinity of a potential barrier, then the resulting electric field fluctuations produce a change in
the scattering amplitude which approximately agrees with that given by the one loop vertex correction to the scattering amplitude. 
This agreement supports the hypothesis that the
details of a given physical situation can define the appropriate switching function.

In the present paper, we are concerned with the coupling of a quantized matter field stress tensor with gravity, specifically with the tensor
perturbation of an expanding universe. This coupling is mediated by Newton's constant, which will be assumed to be time-dependent
in the early universe. This time-dependence produces both of the types of switching discussed above, the switching of the coupling
constant of the theory, and time-averaging of the quantum stress tensor operator. The functional form of the switching and its
physical origin are not  specified, but assumed to arise from some processes in the early universe which are not yet understood.
Several specific choices will be explored in Sec.~\ref{sec:choices}.

\section{free graviton two-point functions and power spectra}
\label{S:free-gravitons}
\subsection{Tensor metric perturbations in spatially flat Robertson-Walker spacetimes}

We start with a brief review of the propagation of gravitational waves in a spatially flat Robertson-Walker universe, 
described by the background metric
\begin{equation}\label{E:RW}
	ds^{2}= -dt^2 +a^2(t)\, \bigl(dx^{2}+dy^{2}+dz^{2}\bigr)
	=a^{2}(\eta)\,\bigl(-d\eta^{2}+dx^{2}+dy^{2}+dz^{2}\bigr)\,,
\end{equation}
where $t$ is comoving time and $\eta$ is conformal time. 
Let the perturbed metric tensor be
\begin{equation}
g_{\mu\nu} = \gamma_{\mu\nu} + h_{\mu\nu}\,,
\label{eq:h-def}
\end{equation}
where the background metric, Eq.~\eqref{E:RW}, is $\gamma_{\mu\nu} $ and $h_{\mu\nu}$ is the perturbation.
We  may impose the transverse, trace-free (TT) gauge conditions
\begin{align}\label{E:TT}
	h^{\mu\nu}{}_{;\,\nu}&=0\,,&h_{\mu}{}^{\mu}&=h=0\,,&h^{\mu\nu}u_{\nu}&=0\,,
\end{align}
which reduce the metric perturbations to two transverse physical degrees of freedom, corresponding to the polarizations of 
the gravitational wave. Here the semicolon denotes the covariant differentiation with respect to the background metric 
$\gamma_{\mu\nu}$, and $u^{\mu}=\delta^{u}_{t}$ is the four-velocity of a comoving observer.

It was shown by Lifshitz~\cite{Lifshitz} that in the absence of sources, the mixed tensor of the metric perturbation 
$h_{\mu}{}^{\nu}$ in this gauge satisfies the scalar wave equation in the spatially flat Robertson-Walker universe,
\begin{equation}\label{E:KG}
	\square\,h_{\mu}{}^{\nu}=0\,,
\end{equation}
where the scalar wave operator $\square$ in the metric Eq.~\eqref{E:RW} takes the form
\begin{equation}
	\square=\frac{1}{\sqrt{-\gamma}}\,\partial_{\mu}\biggl(\sqrt{-\gamma}\,\gamma^{\mu\nu}\partial_{\nu}\biggr)\,,
	\label{E:Sbox}
\end{equation}
with $\gamma$ denoting the determinant of the background metric $\gamma_{\mu\nu}$. This  implies that gravitational 
waves behave like a pair of free minimally coupled massless scalar fields in the the spatially flat Robertson-Walker 
universe~\cite{FordParker:1977}. In general. we may write the solution to Eq.~\eqref{E:KG} as a sum of plane wave modes,
which are of the form
\begin{equation}\label{E:plane-wave}
	\varepsilon_{\mu}{}^{\nu}(\mathbf{k},\lambda)\,{\bf e}^{i\,\mathbf{k}\cdot\mathbf{x}}f_k(\eta)\,,
\end{equation}
where $\mathbf{x}=(x,y,z)$, $\mathbf{k}$ is a wave vector, and $\lambda$ labels the independent polarizations. 
The gauge conditions, Eq.~\eqref{E:TT}, imply that the polarization tensor $\varepsilon_{\mu}{}^{\nu}$ satisfies 
\begin{align}
	k^{\mu}\varepsilon_{\mu}{}^{\nu}&=0\,,&\varepsilon_{\mu}{}^{\mu}&=0\,,&&\text{and}&u^{\mu}\varepsilon_{\mu}{}^{\nu}&=0\,.
\end{align}
When the wavevector $\mathbf{k}$ is in the $z$-direction, the two independent polarization tensors have very simple forms:
\begin{align}
	(+):&\quad\varepsilon_{x}{}^{x}=-\varepsilon_{y}{}^{y}=\frac{1}{\sqrt{2}}\,,&(\times)&:\quad\varepsilon_{x}{}^{y}=
	\varepsilon_{y}{}^{x}=\frac{1}{\sqrt{2}}\,,
\end{align}
while other elements vanish. The temporal part of the mode function, $f_k(\eta)$,  satisfies
\begin{equation}\label{E:mode-eq}
	\partial_{\eta}\Bigl(a^{2}\partial_{\eta}f_k \Bigr)+k^{2}a^{2}\,f_k=0\,,
\end{equation}
and hence depends upon the functional form of the scale factor $a(\eta)$ and upon $k=|\mathbf{k}|$ , 
but is independent of polarization and of the
direction of $\mathbf{k}$. The mode function, $f_k(\eta)$,  is normalized by imposition of the Wronskian condition
\begin{equation}
	f_k(\eta) \partial_{\eta}f^{*}_k(\eta)-f^{*}_k(\eta)\partial_{\eta}f_k(\eta)=\frac{i}{a^{2}(\eta)}\,.
	\label{E:Wronskian}
\end{equation}

With this normalization, we may write the free graviton field on the spatially flat Robertson-Walker background as
\begin{equation}\label{E:grav-field}
\hat{h}^{(1)\;j}_{\;\;\; i}(\eta,\mathbf{x})=2\kappa\int\!\frac{d^{3}k}{(2\pi)^{\frac{3}{2}}}
\sum_{\lambda=1}^{2}\varepsilon_{i}{}^{j}(\mathbf{k},\lambda)\,\hat{a}_{\mathbf{k}\lambda}\,e^{i\,\mathbf{k}\cdot\mathbf{x}}
\,f_k(\eta)+\text{H.c.}
\end{equation}
The graviton creation and annihilation operators $\hat{a}_{\mathbf{k}\lambda}$, and
 $\hat{a}_{\mathbf{k}\lambda}^{\dagger}$ obey
\begin{equation}
	[\hat{a}_{\mathbf{k}\lambda}^{\vphantom{\dagger}},\hat{a}_{\mathbf{k}'\lambda'}^{\dagger}]=
	\delta_{\lambda\lambda'}\delta(\mathbf{k}-\mathbf{k}')\,.
\end{equation}
Here
\begin{equation}
\kappa = \sqrt{8 \pi G_N} = \sqrt{8 \pi}\, \ell_P \,, 
\label{E:kappa}
\end{equation}
where $G_N$ is Newton's constant, $\ell_P$ is the Planck length, and units where $c=\hbar=1$ are adopted.
 The factor of $\kappa$ in 
$\hat{h}^{(1)\;j}_{\;\;\; i}(\eta,\mathbf{x})$ arises because the effective energy momentum tensor for gravitons is
quadratic in $\hat{h}^{(1)\;j}_{\;\;\; i}(\eta,\mathbf{x})$ and proportional to $\kappa^{-2}$. (See the discussion in Sec.~\ref{S:pert}.) 
In addition, we require that the zero point energy of a given graviton mode, which is an integral of the  effective energy momentum 
tensor, have its usual form of one-half of the angular frequency of the mode, which is independent of $\kappa$. 

The above mode expansion defines a vacuum state $|0\rangle$ by $\hat{a}_{\mathbf{k}\lambda}|0\rangle =0$, 
for all $\mathbf{k}$ and $\lambda$. However, this vacuum state is not uniquely defined, as there are an infinite
set of mode functions $f_k(\eta)$ which satisfy both Eqs.~\eqref{E:mode-eq} and \eqref{E:Wronskian}. This is
the usual ambiguity of defining particles in curved spacetime. We will not attempt to address this ambiguity for a 
general scale factor, but will focus on the special case of de Sitter spacetime.

\subsection{Gravitons in de Sitter Spacetime}
 
de Sitter spacetime may be represented as a spatially flat Robertson-Walker universe, Eq.~\eqref{E:RW}, with
\begin{equation}
a(\eta) = -\frac{1}{H\, \eta}\,,
\end{equation}
where $H$ is the Hubble parameter and $-\infty < \eta < 0$. The set of coordinates covers one-half of global
de Sitter spacetime, but this is more than enough for inflationary cosmology, where $\eta$ can be restricted to run
over a finite range. With this choice of scale factor, the solutions of \eqref{E:mode-eq} may be expressed in terms
of Hankel functions as
\begin{equation}
f_k(\eta) = H\left(\frac{\pi}{4}\right)^{\frac{1}{2}}\eta^{3/2}
 \left[c_1  H_{3/2}^{(1)}(k\eta) +c_2 H_{3/2}^{(2)}(k\eta) \right]\,,
\end{equation}
where the Wronskian condition implies
\begin{equation}
|c_2|^2 - |c_1|^2 =1 \,.
\end{equation}
Each allowed choice of $c_1$ and $c_2$, which may be functions of $k$, leads to a different definition of the graviton  
vacuum in de Sitter spacetime. The Bunch-Davies vacuum arises when  $c_2=1$ and $c_1=0$. For 
massive or nonminimally coupled scalar fields, the Bunch-Davies vacuum is also the de Sitter invariant vacuum state.
It is an attractor state in the sense that inflation redshifts the particle content of other states, and causes
them to approach the Bunch-Davies vacuum. However, the Bunch-Davies vacuum does not exist in a strict sense
for massless, minimally coupled scalar field or for the graviton field. This is due to an infrared divergence in the
two-point function. For free gravitons, this function is defined by
\begin{equation}
{}_{11}K_{i}{}^{j}{}_{k}{}^{l}(x,x')=\langle 0 \vert\hat{h}^{(1)\;j}_{\;\;\; i}(x)\hat{h}^{(1)\;l}_{\;\;\; k}(x')\vert0\rangle\,,
\label{E:grav-2pt}
\end{equation}
and contains an integral of the form
\begin{equation}
\int d^3k \, {\bf e}^{i\,\mathbf{k}\cdot(\mathbf{x} - \mathbf{x'} )} \,f_k(\eta)\, f^*_k(\eta')\,.
\label{E:k-int}
\end{equation}
For the  Bunch-Davies vacuum, $f_k(\eta) = u_k(\eta)$, where
\begin{equation}
u_{k}(\eta)=H\left(\frac{\pi}{4}\right)^{\frac{1}{2}}\eta^{3/2}H_{3/2}^{(2)}(k\eta)=
i\,\frac{H}{(2k^{3})^{\frac{1}{2}}}\bigl(1+i\,k\eta\bigr)e^{-i\,k\eta}\,.
\label{E:grav-mode}
\end{equation}
Because $|u_k(\eta)| \propto k^{-3/2}$ as $k \rightarrow 0$, the integral in Eq.~\eqref{E:k-int} diverges logarithmically at its lower
limit. One resolution of this problem is to modify the state slightly for very long wavelength modes~\cite{FP77IR}. This may be done by
allowing $c_1$ and $c_2$ to have the Bunch-Davies values for $k>k_C$, but to vary for $k<k_C$ so as to avoid an
infrared divergence in Eq.~\eqref{E:k-int}. So long as $k_C$ is below any graviton wavenumber with which we are concerned,
this infrared finite state is indistinguishable for the Bunch-Davies vacuum. We will adopt his viewpoint here, and use
Eq.~\eqref{E:grav-mode} as the graviton mode function.

Now the mode expansion of the graviton field operator in de Sitter spacetime becomes
\begin{equation}\label{E:h1}
\hat{h}^{(1)\;j}_{\;\;\; i}(\eta,\mathbf{x})=2\kappa\int\!\frac{d^{3}k}{(2\pi)^{\frac{3}{2}}}
\sum_{\lambda=1}^{2}\varepsilon_{i}{}^{j}(\mathbf{k},\lambda)\,\hat{a}_{\mathbf{k}\lambda}\,e^{i\,\mathbf{k}\cdot\mathbf{x}}
\,u_k(\eta)+\text{H.c.}\,.
\end{equation}

\subsection{Power Spectra}

It is well-known that a power spectrum can be defined as a Fourier transform of a correlation function. In 
the spatially flat Robertson-Walker universe, Eq.~\eqref{E:RW}, let $K(x,x') = K(\mathbf{x} - \mathbf{x'}, \eta,\eta')$
be a field correlation function, or two-point function. Here we assume a state with spatial translation symmetry
in writing  $K$ as a function of $\mathbf{x} - \mathbf{x'}$. Define the power spectrum at time $\eta$ by a spatial
Fourier transform along an equal-time hypersurface, $\eta' = \eta$:
\begin{equation}
P(k,\eta) = \frac{1}{(2\pi)^3} \int d^3u \, {\bf e}^{i\,\mathbf{k}\cdot\mathbf{u} } \,  K(\mathbf{u}, \eta,\eta) \,.
\label{E:power}
\end{equation}
Note that in cosmology, the term ``power spectrum" often refers not to $P(k,\eta)$, but rather to
 \begin{equation}
{\cal P}(k,\eta) =  k^3 \, P(k,\eta) \,.
\end{equation}
(Here we use a definition of ${\cal P}$ which differs by a factor of $4\pi$ from that in Refs.~\cite{FMNWW10,WHFN11},
but which seems to agree with many authors.)

If we adopt a quantum state which is close to the Bunch-Davies vacuum as described above, then the spatial
Fourier transform of Eq.~\eqref{E:grav-2pt} leads to the free graviton power spectum
\begin{equation}\label{E:P11}
	P_{11}(k,\eta)=2\kappa^{2}\!\int\!\frac{d^{3}\mathbf{q}}{(2\pi)^{3}}\;\delta^{(3)}(\mathbf{k}+\mathbf{q})\,
	u_{\mathbf{q},\lambda}^{\vphantom{*}}(\eta)u^{*}_{\mathbf{q},\lambda}(\eta)=
	\frac{\kappa^{2}H^{2}}{4\pi^{3}k^{3}}\bigl(1+k^{2}\eta^{2}\bigr)\,.
\end{equation}
We are often interested in this function evaluated at the end of inflation, $\eta =\eta_r$. In this case, we drop the
explicit $\eta$ dependence in $P$ and write
\begin{equation}
P_{11}(k) = \frac{\kappa^{2}H^{2}}{4\pi^{3}k^{3}}\bigl(1+k^{2}\eta_r^{2}\bigr)\,
\label{E:P11-2}
\end{equation}
or 
\begin{equation}
\mathcal{P}_{11}(k)=\frac{\kappa^{2}H^{2}}{4 \pi^{3}}\bigl(1+k^{2}\eta^{2}_{r}\bigr)\,.
\label{E:P11-3}
\end{equation}
The tensor perturbation coming from free gravitons in de Sitter spacetime were discussed in the context of
inflationary-type models by Starobinsky~\cite{Starobinsky79}, Abbott and Wise~\cite{AW84}, and by Allen~\cite{Allen88},
among others. A recent review of this topic was given by  Guzzetti, {\it et al.}~\cite{GBLM06}.  Note that if 
$k^{2}\eta^{2}_{r} \ll 1$, then Eq.~\eqref{E:P11-3} describes a scale invariant spectrum if $H$ is independent of
$k$. Most inflationary models predict a weak power law dependence, described by the tensor spectral index, which
arises because $H$ is slowly decreasing in time during inflation.

\section{Effects of a Quantum Matter Field}
\label{sec:matter}

So far, we have been concerned with free gravitons propagating on a cosmological background spacetime. Now
we wish to introduce quantized matter, which we take to be a conformal field, such as the electromagnetic field,
with stress tensor operator $\hat{T}_{\mu\nu}$. As is well-known, the expectation value of this operator in any quantum 
state is formally divergent, and needs to be regularized and renormalized by four counterterms  in the Einstein 
equations~\cite{BD}. Two of these
counterterms renormalize the cosmological constant and Newton's constant, respectively.  The second two counterterms
are associated with two counterterms in the gravitational action which are quadratic  in the curvature. These may be
taken to be the square of the scalar curvature and of the Weyl tensor. After renormalization, $\langle\hat{T}_{\mu\nu}\rangle$
may contain contributions from each of the tensors associated with each counterterm. In addition, it will contain a part
with a nonzero trace, the conformal anomaly. If we omit the metric and Einstein tensors from cosmological constant and 
Newton's constant renormalizations, the local geometric part of $\langle\hat{T}_{\mu\nu}\rangle$ on a Robertson-Walker
background may be written as
\begin{equation}\label{E:local}
	\langle\hat{T}_{\mu\nu}[g]\,\rangle_{L}=c_{1}\,\mathcal{H}_{\mu\nu}+c_{2}\,\mathcal{A}_{\mu\nu}+C\,\mathcal{B}_{\mu\nu}\,.
\end{equation}
Here the tensors $\mathcal{H}_{\mu\nu}$, $\mathcal{A}_{\mu\nu}$ and $\mathcal{B}_{\mu\nu}$ are
\begin{align}
	\mathcal{H}_{\mu\nu}&=-2R_{\,;\mu\nu}+2g_{\mu\nu}R_{\,;\,\rho}{}^{\rho}-\frac{1}{2}\,g_{\mu\nu}R^{2}+2RR_{\mu\nu}\,,\\
	\mathcal{A}_{\mu\nu}&=-4\nabla_{\alpha}\nabla_{\beta}C_{\mu}{}^{\alpha}{}_{\nu}{}^{\beta}-2R_{\alpha\beta}
	C_{\mu}{}^{\alpha}{}_{\nu}{}^{\beta}\,,     \label{E:Amunu}  \\
	\mathcal{B}_{\mu\nu}&=-2C_{\alpha\mu\beta\nu}R^{\alpha\beta}+\frac{1}{2}\,g_{\mu\nu}R_{\alpha\beta}R^{\alpha\beta}+
	\frac{2}{3}\,RR_{\mu\nu}-R_{\mu}{}^{\alpha}R_{\nu\alpha}-\frac{1}{4}\,g_{\mu\nu}R^{2}\,,
\end{align}
where $R_{\mu\nu}$ is the Ricci tensor, $R=R^\mu_\mu$ is the scalar curvature, and the Weyl tensor is defined by
\begin{equation}\label{E:Weyl}
	 C_{\alpha\beta\gamma\delta}=R_{\alpha\beta\gamma\delta}-\frac{1}{2}\Bigl[g_{\alpha\gamma}R_{\beta\delta}-
	 g_{\alpha\delta}R_{\beta\gamma}+g_{\beta\delta}R_{\alpha\gamma}-g_{\beta\gamma}R_{\alpha\delta}\Bigr]+
	 \frac{1}{6}\Bigl[g_{\alpha\gamma}g_{\beta\delta}-g_{\alpha\delta}g_{\beta\gamma}\Bigr]R\,,
\end{equation}
for the metric tensor $g_{\alpha\beta}$. The tensor $\mathcal{B}_{\mu\nu}$ will give the trace anomaly of the conformally invariant 
field in the conformally flat spacetime, and its coefficient $C$ takes different values for various conformally invariant fields. 
On the other hand, the coefficients $c_{1}$, $c_{2}$ of the remaining two tensors are undetermined, and are associated with the
two counterterms which are quadratic in the curvature. In addition to
$\langle\hat{T}_{\mu\nu}[g]\,\rangle_{L}$, there can be nonlocal contributions to $\langle\hat{T}_{\mu\nu}\rangle$, which will
be discussed in more detail in Sec.~\ref{S:Qij}.

We will be concerned with the effects of the fluctuations of the matter stress tensor operator, $\hat{T}_{\mu\nu}$, around
its renormalized expectation value, $\langle\hat{T}_{\mu\nu}\rangle$. These stress tensor fluctuations will contribute 
to additional terms in the graviton field, $\hat{h}_{\mu\nu}$ which are higher order  in $\kappa$ than the free graviton field.
However, we first discuss the various expansions which we need.

\subsection{Perturbative expansions}
\label{S:pert}

The treatment of gravity waves on a fixed background necessarily involves an expansion of the metric and other tensors
in powers of the metric perturbation. The perturbed metric tensor was defined in Eq.~\eqref{eq:h-def}.
 We may expand the
Einstein equations in powers of the perturbations, $h_{\mu\nu}$, using the corresponding expansion of the Einstein tensor:
\begin{equation}
G_{\mu\nu} = {}^{(0)}G_{\mu\nu} + {}^{(1)}G_{\mu\nu} +{}^{(2)}G_{\mu\nu} + {}^{(3)}G_{\mu\nu} +\cdots \,. 
\label{E:G-expand}
\end{equation}
In the expansion of the Einstein equations, the lowest order is an equation for the background metric, the first
order gives the equation for linear perturbations, Eq.~\eqref{E:KG}, and the second order term describes the backreaction 
of gravity waves on the background, in the limit where the wavelength of the gravity waves is small
compared to the local radii of curvature of the background geometry~\cite{MTW}.

However, in addition to this expansion, we will need an expansion of the graviton field in powers of $\kappa$,
which acts as the coupling constant for gravity. Write this expansion as
\begin{equation}
\hat{h}_{\mu\nu}=\hat{h}^{(1)}_{\mu\nu}+\hat{h}^{(2)}_{\mu\nu}+\hat{h}^{(3)}_{\mu\nu}+\cdots \,.
\label{E:h-expand}
\end{equation}
This expansion begins at first order, as the free graviton field is already proportional to $\kappa$.
Note that we use a notation in which the order in $h_{\mu\nu}$ is denoted by a superscript on the left,
as in Eq.~\eqref{E:G-expand}, while the order in $\kappa$ is denoted by a superscript on the right,
as in Eq.~\eqref{E:h-expand}.

We will now treat the Einstein equation as an operator relation involving both the graviton field operator,
$\hat{h}_{\mu\nu}$, and the matter stress tensor operator, $\hat{T}_{\mu\nu}$. In the presence of a
cosmological constant $\Lambda$, the Einstein equation is
\begin{equation}\label{E:Einstein}
	G_{\mu\nu}[\gamma+\hat{h}]-\Lambda\bigl(\gamma_{\mu\nu}+\hat{h}_{\mu\nu}\bigr)-
	\kappa^{2}\,\hat{T}_{\mu\nu}[\gamma+\hat{h}]=0\,.
\end{equation}
Next we expand in both $h_{\mu\nu}$ and $\kappa$ to write
 \begin{align}\label{E:expand}
	&\Bigl\{{}^{(0)}G_{\mu\nu}[\gamma]-\Lambda\,\gamma_{\mu\nu}-\kappa^{2}\,\langle{}^{(0)}\hat{T}_{\mu\nu}[\gamma]\,\rangle\Bigr\}
	+\Bigl\{{}^{(1)}G_{\mu\nu}[\hat{h}^{(1)}]-\Lambda\,\hat{h}^{(1)}_{\mu\nu}\Bigr\}\\
	+&\Bigl\{{}^{(1)}G_{\mu\nu}[\hat{h}^{(2)}]-\Lambda\,\hat{h}^{(2)}_{\mu\nu}-\kappa^{2}\Bigl({}^{(0)}\hat{T}_{\mu\nu}[\gamma]
	-\langle{}^{(0)}\hat{T}_{\mu\nu}[\gamma]\,\rangle\Bigr)\Bigr\}\notag\\
	+&\Bigl\{{}^{(1)}G_{\mu\nu}[\hat{h}^{(3)}]-\Lambda\,\hat{h}^{(3)}_{\mu\nu}-
	\kappa^{2}\,{}^{(1)}\hat{T}_{\mu\nu}[\hat{h}^{(1)}]\Bigr\}+
	{}^{(2)}G_{\mu\nu}[\hat{h}^{(1)}]+{}^{(3)}G_{\mu\nu}[\hat{h}^{(1)}]+\mathcal{O}(\kappa^{4})=0\,.\notag
\end{align}

The first term in braces is  zero-th order in the metric perturbation, and its vanishing is the equation
for the background with the expectation value $\langle{}^{(0)}\hat{T}_{\mu\nu}[\gamma]\rangle$ as its source,
\begin{equation}
{}^{(0)}G_{\mu\nu}[\gamma]-\Lambda\,\gamma_{\mu\nu} =
\kappa^{2}\,\langle{}^{(0)}\hat{T}_{\mu\nu}[\gamma]\rangle \,.
\label{E:h0-eq}
\end{equation}
In de Sitter spacetime, $\langle{}^{(0)}\hat{T}_{\mu\nu}[\gamma]\rangle \propto \gamma_{\mu\nu}$ 
and is hence a finite
shift in the value of the cosmological constant. The second term in braces is first order both in  $h_{\mu\nu}$ and 
in $\kappa$, and its vanishing is the equation for the perturbation. Of the remaining terms, there are two which
do not depend upon the conformal stress tensor. At the classical level, ${}^{(2)}G_{\mu\nu}$ describes the 
backreaction of gravity waves on the background geometry, as noted above. At the quantum level, both 
${}^{(2)}G_{\mu\nu}[\hat{h}^{(1)}]$ and ${}^{(3)}G_{\mu\nu}[\hat{h}^{(1)}]$ describe graviton loop effects,
which we will ignore here.

The third and fourth terms in braces in Eq.~\eqref{E:expand} are of orders $\kappa^2$ and $\kappa^3$, 
respectively, and may be separately set to zero, leading to 
\begin{equation}
{}^{(1)}G_{\mu\nu}[\hat{h}^{(2)}]-\Lambda\,\hat{h}^{(2)}_{\mu\nu}= \kappa^{2}\Bigl({}^{(0)}\hat{T}_{\mu\nu}[\gamma]-\langle{}^{(0)}\hat{T}_{\mu\nu}[\gamma]\,\rangle\Bigr)\,,
\label{E:h2-eq}
\end{equation}
and
\begin{equation}
{}^{(1)}G_{\mu\nu}[\hat{h}^{(3)}]-\Lambda\,\hat{h}^{(3)}_{\mu\nu}=\kappa^{2}\,{}^{(1)}\hat{T}_{\mu\nu}[\hat{h}^{(1)}]\,.
\label{E:h3-eq}
\end{equation}
Equation~\eqref{E:h2-eq} relates $\hat{h}^{(2)}_{\mu\nu}$ to the fluctuations of the matter stress tensor, and
Eq.~\eqref{E:h3-eq} relates $\hat{h}^{(3)}_{\mu\nu}$ to ${}^{(1)}\hat{T}_{\mu\nu}[\hat{h}^{(1)}]$, the shift in the matter
stress tensor operator due to a first order perturbation of the background. The solutions of these equations will
be discussed in Secs.~\ref{S:P22}   and  \ref{S:P13}, but here we note that both may be simplified in the transverse, trace-free gauge.
We assume that the TT gauge conditions, Eq.~\eqref{E:TT}, apply in each order of the expansion, 
Eq.~\eqref{E:h-expand}. The equation for the first order term in this gauge may be written as
\begin{equation}
{}^{(1)}G_{\mu}{}^{\nu}[\hat{h}^{(1)}]-\Lambda\,\hat{h}^{(1)}_{\mu}{}^{\nu} =
-\frac{1}{2} \square\, \hat{h}^{(1)}_{\mu}{}^{\nu}=0\,,
\label{E:h1-eq}
\end{equation}
which is the Lifshitz equation, Eq.~\eqref{E:KG}. Note that the  left-hand sides of Eqs.~\eqref{E:h2-eq}, \eqref{E:h3-eq}, and
 \eqref{E:h1-eq} have the same functional form. Thus we may express the
former two equations as
\begin{equation}
\square\,\hat{h }^{(2)}_{\mu\nu} = 
-2 \kappa^{2}\Bigl({}^{(0)}\hat{T}_{\mu\nu}[\gamma]-\langle{}^{(0)}\hat{T}_{\mu\nu}[\gamma]\,\rangle\Bigr)\,,
\label{E:h2-eq2}
\end{equation}
and
\begin{equation}
\square\, \hat{h}^{(3)}_{\mu\nu} = -2\kappa^{2}\,{}^{(1)}\hat{T}_{\mu\nu}[\hat{h}^{(1)}]\,,
\label{E:h3-eq2}
\end{equation}
where $\square$ is the scalar wave operator, defined in Eq.~\eqref{E:Sbox}.
In the following sections, we will use these two equations to calculate the $O(\kappa^4)$ corrections
to the gravity wave power spectrum.

\subsection{Graviton Correlation Functions}
\label{S:corr-fnts}

Let $|\Psi\rangle$ denote the quantum state of the combined graviton-conformal field system. 
As discussed earlier, we take the gravitons to be in an approximation to the Bunch-Davies vacuum,
in which Eq.~\eqref{E:grav-mode} is the graviton mode function, and denote this state by
$\vert0_{G}\rangle$. The matter field is assumed to be in the conformal vacuum state, $\vert0_{M}\rangle$,
in which the matter field correlation functions are conformal transforms of the Minkowski space
vacuum correlation functions. We assume that the state of the combined system may be written
as a direct product: $ |\Psi\rangle = \vert0_{G}\rangle \, \vert0_{M}\rangle$.

The full graviton corrections function
\begin{equation}
K_{\mu\nu\rho\sigma}(x,x')=\langle\Psi\vert\hat{h}_{\mu\nu}(x)\hat{h}_{\rho\sigma}(x')\vert\Psi\rangle\,,
\end{equation}
may be expanded in powers of $\kappa$ using Eq.~\eqref{E:h-expand} as
\begin{equation}
K_{\mu\nu\rho\sigma}(x,x')= {}_{11}K_{\mu\nu\rho\sigma}(x,x') + {}_{22}K_{\mu\nu\rho\sigma}(x,x') 
+ {}_{13}K_{\mu\nu\rho\sigma}(x,x') +\cdots \,.
\end{equation}
Here
\begin{equation}
 {}_{11}K_{\mu\nu\rho\sigma}(x,x') = 
 \langle\Psi\vert\hat{h}^{(1)}_{\mu\nu}(x)\hat{h}^{(1)}_{\rho\sigma}(x')\vert\Psi\rangle =
 \langle 0_G\vert\hat{h}^{(1)}_{\mu\nu}(x)\hat{h}^{(1)}_{\rho\sigma}(x')\vert 0_G \rangle 
\end{equation}
is the free graviton correlation function defined in Eq.~\eqref{E:grav-2pt}. It is independent of the matter
state, $\vert0_{M}\rangle$, and of order $\kappa^2$. There is no order $\kappa^3$ contribution, as
$\langle 0_G\vert\hat{h}^{(1)}_{\mu\nu}(x)\vert 0_G \rangle =0$. There are two  order $\kappa^4$ terms, 
\begin{equation}
{}_{22}K_{\mu\nu\rho\sigma}(x,x') = 
 \langle\Psi\vert\hat{h}^{(2)}_{\mu\nu}(x)\hat{h}^{(2)}_{\rho\sigma}(x')\vert\Psi\rangle\,,
 \label{E:K22}
\end{equation}
and
\begin{equation}
{}_{13}K_{\mu\nu\rho\sigma}(x,x') = 
\frac{1}{2}\, \left[\langle\Psi\vert\hat{h}^{(1)}_{\mu\nu}(x)\hat{h}^{(3)}_{\rho\sigma}(x')\vert\Psi\rangle+
\langle\Psi\vert\hat{h}^{(3)}_{\mu\nu}(x)\hat{h}^{(1)}_{\rho\sigma}(x')\vert\Psi\rangle \right]\,.
 \label{E:K13}
\end{equation}
These will be studied in Secs.~\ref{S:P22} and \ref{S:P13}, respectively.

\section{Gravity Waves from Stress Tensor Fluctuations}
\label{S:P22}

In this section, we deal with the $O(\kappa^2)$ part of the graviton field, $\hat{h }^{(2)}_{\mu}{}^{\nu}$, 
 and its contribution to the  power spectrum. Equation~\eqref{E:h2-eq2} reveals that 
$\hat{h }^{(2)}_{\mu}{}^{\nu}$ describes gravity waves radiated by the fluctuations of the conformal
stress tensor. We impose the initial condition that $\hat{h }^{(2)}_{\mu}{}^{\nu}\rightarrow 0$ if
the right-hand side of Eq.~\eqref{E:h2-eq2} vanishes, which assumes no incoming radiation from
other sources. Then the solution becomes
 \begin{equation}\label{E:h2soln}
	\hat{h}^{(2)}_{\mu\nu}=2\kappa^{2}\!\int\!d^{4}x'\sqrt{-\gamma(x')}\;G_{R}(x,x')\Bigl({}^{(0)}\hat{T}_{\mu\nu}[\gamma]
	-\langle{}^{(0)}\hat{T}_{\mu\nu}[\gamma]\,\rangle\Bigr)\,,
\end{equation}
where $G_{R}(x,x')$ is the retarded Green's function for the scalar wave operator, which satisfies
\begin{equation}\label{E:GR}
	\square_{x}\,G_{R}(x,x')=-\frac{\delta^{(4)}(x-x')}{\sqrt{-\gamma'}}\,.
\end{equation}
The spatial Fourier transform of $G_{R}(x,x')$  is (See, for example, Refs.~\cite{WKF07,WHFN11}.) 
\begin{equation}\label{E:Gk}
	\widetilde{G}_{R}(\eta,\eta'; {k})=\frac{H^{2}}{(2\pi k)^{3}}\,
	\left[\bigl(1+k^{2}\eta\eta'\bigr)\sin k(\eta-\eta')-k(\eta-\eta')\cos k(\eta-\eta')\right]\,,
\end{equation}
in de Sitter spacetime. Recall that our normalization for Fourier transforms is given in Eq.~\eqref{E:power}.

The correlation function associated with $\hat{h }^{(2)}_{\mu}{}^{\nu}$ was defined in Eq.~\eqref{E:K22}. 
However, the operator  $\hat{h }^{(2)}_{\mu}{}^{\nu}$ acts only on the matter field, so
\begin{equation}
{}_{22}K_{i}{}^{j}{}_{k}{}^{l}(x,x')=\langle0_{M}\vert\hat{h}^{(2)\;j}_{\;\;\; i}(x)\hat{h}^{(2)\;l}_{\;\;\; k}(x')\vert0_{M}\rangle\,,
\label{E:K22-1}
\end{equation}
which may be written as
\begin{align}\label{E:K22-2}
	&\quad{}_{22}K_{i}{}^{j}{}_{k}{}^{l}(x,x')\notag\\
	&=\bigl(2\kappa^{2}\bigr)^{2}\!\int\!d^{4}x_{1}\sqrt{-\gamma(x_{1})}\!\int\!d^{4}x_{2}\sqrt{-\gamma(x_{2})}\;
	G_{R}(x,x_{1})G_{R}(x',x_{2})\,C_{i}{}^{j}{}_{k}{}^{l}(x_{1},x_{2})\,.
\end{align}
Here $C_{i}{}^{j}{}_{k}{}^{l}(x_{1},x_{2})$ is the correlation function for the stress tensor, defined by
\begin{equation}
	C_{\mu}{}^{\nu}{}_{\rho}{}^{\sigma}(x,x')=\langle0_{M}\vert{}^{(0)}\hat{T}_{\mu}{}^{\nu}(x){}^{(0)}\hat{T}_{\rho}{}^{\sigma}(x')\vert0_{M}
	\rangle-\langle0_{M}\vert{}^{(0)}\hat{T}_{\mu}{}^{\nu}(x)\vert0_{M}\rangle\langle0_{M}
	\vert{}^{(0)}\hat{T}_{\rho}{}^{\sigma}(x')\vert0_{M}\rangle\,.
\end{equation}
More precisely, if ${}_{22}K_{i}{}^{j}{}_{k}{}^{l}(x,x')$ is expressed in the transverse, trace-free gauge, then
$C_{i}{}^{j}{}_{k}{}^{l}(x_{1},x_{2})$ is the transverse, trace-free part of the full stress tensor correlation function,
which is best defined in momentum space, as discussed in Ref.~\cite{WHFN11}.

We now implement the switching of Newton's constant discussed in Sec.~\ref{sec:switch}, and let 
$\kappa\to\kappa\,g(\eta)$, so that Newton's constant becomes proportional to $s(\eta) = g^2(\eta)$. We
require that the switching function, $g(\eta)$ satisfy the following conditions: (1) $g(\eta)$ and at least its first
four derivative be finite everywhere; (2) $g(\eta) \leq 1$ everywhere; (3) $g(\eta_r) = 1$; (4) the characteristic interval
in conformal time during which $g\not=0$ be $|\eta_0|$, the approximate duration of inflation;
and (5) $g(\eta) \rightarrow 0$
as $\eta \rightarrow -\infty$ sufficiently rapidly that all integrals on $\eta$ converge at their lower limits.   Concerning (1), we should comment
that if we wish to calculate all of the moments of the stress tensor, as needed for the complete probability distribution, then we must
require that $g(\eta)$ be $C^\infty$. Here we are solely concerned with the variance, so $C^4$ is sufficient..

We can rewrite Eq.~\eqref{E:K22-2} as
\begin{align}
	&\quad{}_{22}K_{i}{}^{j}{}_{k}{}^{l}(x,x') \nonumber \\
	&=\bigl(2\kappa^{2}\bigr)^{2}\int\!d^{4}x_{1}\sqrt{-\gamma(x_{1})}\int\!d^{4}x_{2}\sqrt{-\gamma(x_{2})}\;
	g^{2}(\eta_{1})G_{R}(x,x_{1})\,g^{2}(\eta_{2})G_{R}(x',x_{2})\,C_{\mu}{}^{\nu}{}_{\rho}{}^{\sigma}(x_{1},x_{2})\,,
\label{E:K22-3}	
\end{align}
where now the integrations on both $\eta_1$ and $\eta_2$ range from $-\infty$ to $\eta_r$, the value of the conformal
time on the reheating surface. Note that the factors of $s(\eta) = g^2(\eta)$ act to give time averages of the stress
tensor operators in the sense described in Eq.~\eqref{E:aver}.
 The stress tensor correlation function in an expanding universe may be obtained
from  that in  Minkowski spacetime, $C^{(M)\;\nu}_{\quad\mu}{}_{\rho}{}^{\sigma}(x_{1},x_{2})$,
by a conformal transformation,
\begin{equation}\label{E:conformal}
	C_{\mu}{}^{\nu}{}_{\rho}{}^{\sigma}(x_{1},x_{2})=
	a^{-4}(\eta_{1})\,C^{(M)\;\nu}_{\quad\mu}{}_{\rho}{}^{\sigma}(x_{1},x_{2})\,a^{-4}(\eta_{2})\,.
\end{equation}
Note that the conformal symmetry is broken by the conformal anomaly, the appearance of a nonzero trace of
$\langle0_{M}\vert{}^{(0)}\hat{T}_{\mu\nu}(x)\vert0_{M}\rangle$. However, the conformal
anomaly contribution cancels in the stress tensor correlation function, so we may still use Eq.~\eqref{E:conformal}.

We next take a spatial Fourier transform of Eq.~\eqref{E:K22-3}. The Fourier transform of the right-hand side may
be expressed as an integral of products of the Fourier transformed retarded Green's function, 
 $\tilde{G}_{R}(\eta,\eta'; {k})$, and stress tensor correlation function, 
 $\widetilde{C}_{\mu}{}^{\nu}{}_{\rho}{}^{\sigma}(\eta_{1},\eta_{2};\mathbf{k})$. The result is $P_{22}({k})$,
 the contribution to the power spectrum from stress tensor fluctuations:
\begin{align}\label{E:P22-0}
	P_{22}({k})&=\int\!\frac{d^{3}\mathbf{R}}{(2\pi)^{3}}\;{}_{22}K_{i}{}^{j}{}_{k}{}^{l}(x,x')\,
	{\rm e}^{i\,\mathbf{k}\cdot\mathbf{R}}\,\Big|_{\eta=\eta'=\eta_{r}}\\
	&=\bigl(2\kappa^{2}\bigr)^{2}(2\pi)^{6}\!\int_{-\infty}^{\eta_{r}}\!d\eta_{1}\!\int_{-\infty}^{\eta_{r}}\!d\eta_{2}\;g^{2}(\eta_{1}) 
	\widetilde{G}_{R}(\eta_{r},\eta_{1};{k})\,g^{2}(\eta_{2})\widetilde{G}_{R}(\eta_{r},\eta_{2};{k})\,
	\widetilde{C}_{\mu}{}^{\nu}{}_{\rho}{}^{\sigma}(\eta_{1},\eta_{2};\mathbf{k})\,.\notag
\end{align}     

In the transverse, trace-free gauge, when the wave vector points along the $z$-axis, the relevant components of the 
stress-tensor correlation function will be $\widetilde{C}_{x}{}^{y}{}_{x}{}^{y}$, $\widetilde{C}_{x}{}^{x}{}_{x}{}^{x}$, 
and the terms obtained by permutations of the indices. In this case, these correlation functions are equal, that is, 
 $\widetilde{C}\equiv\widetilde{C}_{\times}=\widetilde{C}_{+}$, so we drop the polarization label and write the stress 
 tensor correlation function in Minkowski spacetime, appearing in Eq.~\eqref{E:P22-0}, as
\begin{align}\label{E:C}
	\widetilde{C}^{(M)}(\eta,\eta';{k})&=\frac{1}{1280\pi^{5}}\left(\frac{d^{2}}{d\tau^{2}}+
	k^{2}\right)^{2}\left[-\frac{\sin k\tau}{\tau}+\pi\,\delta(\tau)\right]\,,
\end{align}
with $\tau=\eta-\eta'$. The derivation of this result is given in the Appendix.
Next we consider separately the contributions of the two terms in brackets in the above
expression.

\subsection{Delta Function Term}

The contribution of the $\delta(\tau)$ term in Eq.~\eqref{E:C} to $P_{22}$ is proportional to
\begin{align}
	I&=\int_{-\infty}^{\eta_{r}}\!d\eta_{1}\!\int_{-\infty}^{\eta_{r}}\!d\eta_{2}\;g^{2}(\eta_{1})
	\widetilde{G}_{R}(\eta_r,\eta_{1};{k})\,g^{2}(\eta_{2})\widetilde{G}_{R}(\eta_r,\eta_{2};{k})\notag\\
	&\qquad\qquad\qquad\qquad\qquad\qquad\times\left(\frac{d^{2}}{d\eta_{1}^{2}}+k^{2}\right)
	\left(\frac{d^{2}}{d\eta_{2}^{2}}+k^{2}\right)\,\pi\,\delta(\eta_{1}-\eta_{2})\,.\label{E:I}
\end{align}
By the change of the variables from $\eta_{1}$, $\eta_{2}$ to $u=\eta_{1}+\eta_{2}$ and $v=\eta_{1}-\eta_{2}$ and with 
the help of integration by parts, we can write $I$ in the form
\begin{align}
	I&=\frac{\pi}{2}\int^{2\eta_{r}}_{-\infty}\!du\!\int_{u-2\eta_{r}}^{2\eta_{r}-u}\!dv\,\delta(v)\\
	&\qquad\times\left(\frac{d^{2}}{dv^{2}}+k^{2}\right)^{2}\biggl[g^{2}\left(\frac{u+v}{2}\right)
	\widetilde{G}_{R}\left(\eta_{r},\frac{u+v}{2};{k}\right)\,g^{2}\left(\frac{u-v}{2}\right)
	\widetilde{G}_{R}\left(\eta_{r},\frac{u-v}{2};{k}\right)\biggr]\,\,.\notag
\end{align}
There is no surface term because $\delta^{(n)}(v)=0$ when $v\neq0$. 
When the switching function $g$ varies much more slowly than does the retarded Green's function $\widetilde{G}_{R}$  
with respect to $u$ and $v$, we can pull the switching function outside the differential operators, arriving at
\begin{align}\label{E:I2}
	I&\simeq\frac{\pi}{2}\int^{2\eta_{r}}_{-\infty}\!du\!\int_{u-2\eta_{r}}^{2\eta_{r}-u}\!dv\;\delta(v)\,g^{2}\left(\frac{u+v}{2}\right)\,
	g^{2}\left(\frac{u-v}{2}\right)   \\
	&\qquad\qquad\qquad\qquad\qquad\qquad\times\left(\frac{d^{2}}{dv^{2}}+k^{2}\right)^{2}\biggl[\widetilde{G}_{R}\left(\eta_{r},
	\frac{u+v}{2};{k}\right)\,\widetilde{G}_{R}\left(\eta_{r},\frac{u-v}{2};{k}\right)\biggr]\,.\notag
\end{align}
If we perform first the differentiations and then the $v$-integration, and drop terms which oscillate rapidly in the remaining
integral, then we have
\begin{equation}
I \approx \pi\,k^{4}\bigl(1+k^{2}\eta^{2}_{r}\bigr)\,\frac{H^{4}}{(2\pi k)^{6}}\int_{-\infty}^{\eta_{r}}\!d\eta'\;g^{4}(\eta')\,.
\label{E:I3}
\end{equation}
We are primarily interested in contributions which grow as the switching interval increase, but terms which oscillate on a scale of order
$1/k$ will remain constant, and hence may be ignored.

\subsection{Remaining Contribution to $P_{22}$}

The contribution to $P_{22}$, coming from the  $\sin\, k\tau$ term in Eq.~\eqref{E:C}, 
is proportional to the integral,
\begin{align}\label{E:J}
	J&=-\int_{-\infty}^{\eta_{r}}\!d\eta_{1}\!\int_{-\infty}^{\eta_{r}}\!d\eta_{2}\;g^{2}(\eta_{1})\,
	\widetilde{G}_{R}(\eta_{r},\eta_{1};{k})\,g^{2}(\eta_{2})\,\widetilde{G}_{R}(\eta_{r},\eta_{2};{k})\notag\\
	&\qquad\qquad\qquad\qquad\qquad\qquad\times\left(\frac{d^{2}}{d\eta_{1}^{2}}+k^{2}\right)\left(\frac{d^{2}}{d\eta_{2}^{2}}
	+k^{2}\right)\,\frac{\sin k(\eta_{1}-\eta_{2})}{\eta_{1}-\eta_{2}}\,.
\end{align} 
Note that with this definition, $P_{22}$ may be expressed as
\begin{equation}
P_{22} = \frac{1}{5} \, \pi \, \kappa^2\, (I + J).
\label{E:IJ}
\end{equation}
We again change the variables to $u$ and $v$, and introduce the function $N(v)$
\begin{align}
	N(v)&\equiv\left(\frac{d^{2}}{d\eta_{1}^{2}}+k^{2}\right)\left(\frac{d^{2}}{d\eta_{2}^{2}}+k^{2}\right)\,
	\frac{\sin k(\eta_{1}-\eta_{2})}{\eta_{1}-\eta_{2}}=\left(\frac{d^{2}}{dv^{2}}+k^{2}\right)^{2}\frac{\sin kv}{v}\,.
\end{align}    
We note that $N(v)$ is sharply peaked at $v=0$, with $N(0)\sim\mathcal{O}(k^{5})$, within an interval 
$\lvert v\rvert\sim\mathcal{O}(k^{-1})$, and then it falls off to zero very rapidly when $v\gg k^{-1}$. On the other hand, 
the product $\widetilde{G}_{R}(\eta_{1},\eta_{1};{k})\widetilde{G}_{R}(\eta_{r},\eta_{2};{k})$ is regular and 
finite, but oscillates very fast with respect to $u$ and $v$ for large values of $k$. This implies that the bounds 
of the integral over $v$ in Eq.~\eqref{E:J} can be extended from $u-2\eta_r \leq v \leq 2\eta_r -u$ to $-\infty < v < \infty$. 
Since the switching function barely changes within the central peak of $N(v)$, we can move the switching function 
$g(\frac{u\pm v}{2})$ out of the $v$-integral and evaluate it at $v=0$. This will give
\begin{align}
	J& =-\frac{1}{2}\int_{-\infty}^{2\eta_{r}}\!du\!\int_{u-2\eta_{r}}^{2\eta_{r}-u}\!dv\, g^{2}\left(\frac{u+v}{2}\right)g^{2}\left(\frac{u-v}{2}\right)\,
	\widetilde{G}_{R}\left(\eta_{r},\frac{u+v}{2};{k}\right)\widetilde{G}_{R}\left(\eta_{r},\frac{u-v}{2};{k}\right) N(v)\notag\\
	&\approx-\frac{1}{2}\int_{-\infty}^{2\eta_{r}}\!du\;g^{4}\left(\frac{u}{2}\right)\int_{-\infty}^{\infty}\!dv\;
	\widetilde{G}_{R}\left(\eta_{r},\frac{u+v}{2};{k}\right)\widetilde{G}_{R}\left(\eta_{r},\frac{u-v}{2};{k}\right)\,N(v)\,.
\end{align}    
 This approximation introduces an error of order ${O}(\eta_{r})$, which results from the regime 
 $\lvert u-2\eta_{r}\rvert\lesssim\mathcal{O}(k^{-1})$. It is negligible compared to the dominant contribution 
 to the integral for the case $\lvert\eta_{r}/\eta_{0}\rvert\ll1$, where $|\eta_0|$ is the approximate duration of
 inflation in conformal time. 
   
 The product $G_{R}(\eta,\eta_{1};{k})G_{R}(\eta,\eta_{2};{k})$ contains various terms that rapidly 
 oscillate in $u$, $v$, so after carrying out the integral over $v$, we only keep terms that grow in $u$. Hence we find 
 that $J$ becomes
\begin{align}\label{E:J1}
	J&\approx-\frac{k^{4}\pi}{2}\,\bigl(1+k^{2}\eta^{2}_{r}\bigr)\,\frac{H^{4}}{(2\pi k)^{6}}\int_{-\infty}^{\eta_{r}}\!
	d\eta \; g^{4}(\eta)\,.
\end{align}
This result is half of Eq.~\eqref{E:I3} and it takes a minus sign. 
Thus the $P_{22}$ component of the $\mathcal{O}(\kappa^{4})$ contribution to the graviton power spectrum
becomes, using Eqs.~\eqref{E:kappa} and \eqref{E:IJ},
\begin{align}
	P_{22}(\mathbf{k})&=\frac{1}{10\pi^{2}}\frac{\ell_P^4 \, H^{4}}{k^{2}}\,\bigl(1+k^{2}\eta_{r}^{2}\bigr)
	\int_{-\infty}^{\eta_{r}}\!d\eta \;g^{4}(\eta)\,,\label{E:P22}
\end{align}
which is strictly positive and $\propto |\eta_0|$. If the correlation functions are nonsingular, then the positivity of the power spectrum is a
consequence of the Wiener-Khinchin theorem. However, in quantum field theory, with singular correlation functions,
this conclusion does not necessarily follow, as was discussed in Ref.~\cite{HWF10}.

 \section{ The $P_{13}$ contribution to the power spectrum}
 \label{S:P13}
 
 In addition to $P_{22}$, which was computed in the previous section, there is another $\mathcal{O}(\kappa^{4})$ 
 contribution to the power spectrum, $P_{13}$.   This contribution is the Fourier transform of 
 ${}_{13}K_{i}{}^{j}{}_{k}{}^{l}(x,x')$, defined in Eq.~\eqref{E:K13}. It is the cross term in the graviton correlation
 function between the free graviton field,  $\hat{h}^{(1)}_{\mu\nu}$, and $\hat{h}^{(3)}_{\mu\nu}$, which satisfies
 Eq.~\eqref{E:h3-eq2}. We can view $\hat{h}^{(3)}_{\mu\nu}$ as describing the gravitons radiated by the perturbed
 stress tensor, ${}^{(1)}\hat{T}_{\mu\nu}[\hat{h}^{(1)}]$. However, both of these quantities are operators acting on both
the graviton and matter vacuum states.  The process of forming   ${}_{13}K_{i}{}^{j}{}_{k}{}^{l}(x,x')$ will involve
taking an expectation value in both vacua. Because the free graviton field, $\hat{h}^{(1)}_{\mu\nu}$, does not
act on the matter vacuum, $\vert 0_{M}\rangle$, we may take the expectation value of Eq.~\eqref{E:h3-eq2} in this state,
and write a solution of the resulting equation as
\begin{equation}
\langle \hat{h}^{(3)}_{\mu\nu} \rangle_M = 2\kappa^{2}\!\int\!d^{4}x'\sqrt{-\gamma(x')}\;G_{R}(x,x')\, g^2(\eta')\,
\langle   {}^{(1)}\hat{T}_{\mu\nu}[\hat{h}^{(1)}] \rangle_M \,, 
\label{E:h3-1}
\end{equation}
 where we have assumed no incoming solution of the homogeneous equation, and use a notation where $\langle \; \rangle_M$
 denotes an expectation value in the state $\vert0_{M}\rangle$. We have also introduced a factor of $g^2(\eta')$ to describe
 the switch-on of Newton's constant.
 
 Note that $\langle \hat{h}^{(3)}_{\mu\nu} \rangle_M$ is still an operator in the graviton state space, and may be expressed in
 terms of graviton creation and annihilation operators in an expansion analogous to that for the free graviton field, 
 Eq.~\eqref{E:h1},
\begin{equation}\label{E:h3}
\langle\hat{h}^{(3)\;j}_{\;\;\; i}(\eta,\mathbf{x})\rangle_M=2\kappa\int\!\frac{d^{3}k}{(2\pi)^{\frac{3}{2}}}
\sum_{\lambda=1}^{2}\varepsilon_{i}{}^{j}(\mathbf{k},\lambda)\,\hat{a}_{\mathbf{k}\lambda}\,e^{i\,\mathbf{k}\cdot\mathbf{x}}
\,z_k(\eta)+\text{H.c.}
\end{equation}
There is a similar expansion for $\langle   {}^{(1)}\hat{T}_{\mu\nu}[\hat{h}^{(1)}] \rangle_M$, which is an operator in the
graviton state space:
 \begin{equation}\label{E:T1}
\langle   {}^{(1)}\hat{T}_{\mu\nu}[\hat{h}^{(1)}] \rangle_M = 2\kappa\int\!\frac{d^{3}k}{(2\pi)^{\frac{3}{2}}}
\sum_{\lambda=1}^{2}\varepsilon_{i}{}^{j}(\mathbf{k},\lambda)\,\hat{a}_{\mathbf{k}\lambda}\,e^{i\,\mathbf{k}\cdot\mathbf{x}}
\,v_k(\eta)+\text{H.c.}
\end{equation}

 We may now construct a compact expression for $P_{13}(k)$ by inserting the mode expansions in Eqs.~\eqref{E:h1} and
 \eqref{E:h3} into the expression for ${}_{13}K_{i}{}^{j}{}_{k}{}^{l}(x,x')$, Eq.~\eqref{E:K13}, and then taking a Fourier
 transform to write
 \begin{equation}
P_{13}(k) = \frac{\kappa^2}{2 \pi^2}\; {\rm Re} \left[ z_k(\eta_r)\, u^*_k(\eta_r) \right]\,.
\label{E:P13}
\end{equation}
Note that the mode functions are evaluated at the end of inflation, $\eta = \eta_r$. 
Thus the mode function $z_k(\eta_r)$ may be expressed as
\begin{equation}\label{E:zk}
	z_k(\eta_r) = 2\kappa^{2}(2\pi)^{3}  \int^{\eta_r}_{-\infty} d\eta \;a^{4}(\eta) \,\widetilde{G}_{R}(\eta_r,\eta;{k})\,
	g^{2}(\eta) \, v_{k}(\eta)\,,
\end{equation}
 where  $\widetilde{G}_{R}(\eta_r,\eta;{k})$ is defined in Eq.~\eqref{E:Gk}.
 
 Although $\langle   {}^{(1)}\hat{T}_{\mu\nu}[\hat{h}^{(1)}] \rangle_M$ is an operator in the graviton state space, it may be calculated
 as  the expectation value of a conformal field stress tensor in an almost conformally flat spacetime. A formalism for
 this calculation was developed by Horowitz and Wald~\cite{HW82}, and will be briefly reviewed here. See Ref.~\cite{HFLY10}
 for more details, especially concerning the application of the Horowitz-Wald formalism to gravity waves in de Sitter
 spacetime. 
 The explicit expression for $\langle{}^{(1)}\hat{T}_{\mu\nu}[\hat{h}^{(1)}]\,\rangle_M$ contains a sum of four tensors,
\begin{equation}\label{E:T1a}
	\langle{}^{(1)}\hat{T}_{\mu\nu}[\hat{h}^{(1)}]\,\rangle_M=c_{1}\,{}^{(1)}\mathcal{H}_{\mu\nu}[\hat{h}^{(1)}]+
	C\;{}^{(1)}\mathcal{B}_{\mu\nu}[\hat{h}^{(1)}]+\mathcal{P}_{\mu\nu}[\hat{h}^{(1)}]+\mathcal{Q}_{\mu\nu}[\hat{h}^{(1)}]\,.
\end{equation}
 The first two tensors, ${}^{(1)}\mathcal{H}_{\mu\nu}$, ${}^{(1)}\mathcal{B}_{\mu\nu}$ are first-order corrections of the local 
 geometric tensors $\mathcal{H}_{\mu\nu}$ and $\mathcal{B}_{\mu\nu}$ due to the metric perturbation $\hat{h}^{(1)}_{\mu\nu}$.
 Recall that $\mathcal{H}_{\mu\nu}$ arises from an $R^2$ term in the gravitational action, so we absorb this tensor into
 a renormalization of the coefficient of this term, and set $c_1=0$ for our purposes. The tensor ${}^{(1)}\mathcal{B}_{\mu\nu}$
 is the first order perturbation of the conformal anomaly. It was shown in Sec.~III of  Ref.~\cite{HFLY10} that this term may be
 absorbed in a combination of cosmological constant and Newton's constant renormalization, so we will not consider it further. 
  
  The two remaining terms in $\langle{}^{(1)}\hat{T}_{\mu\nu}[\hat{h}^{(1)}]\,\rangle_M$ are 
 \begin{equation}
 \mathcal{P}_{\mu\nu}=
 -16\pi\alpha\,a^{-2}\tilde{\partial}^{\rho}\tilde{\partial}^{\sigma}\left[ {}^{(1)}\tilde{C}_{\mu\rho\nu\sigma}\ln(a)\right]\,,
 \label{E:Pmunu}
\end{equation}
and
\begin{equation}
\mathcal{Q}_{\mu\nu} =\alpha\,a^{-2}\int\!d^{4}x'\;H_{\lambda}(x-x')\,{}^{(1)}\!\tilde{\mathcal{A}}_{\mu\nu}(x')\,.
 \label{E:Qmunu}
\end{equation}
Here ${}^{(1)}\tilde{C}_{\mu\rho\nu\sigma}$ is the Weyl tensor of perturbed Minkowski spacetime with the perturbation 
$\tilde{h}_{\mu\nu}=a^{-2}h^{(1)}_{\mu\nu}$,  or equivalently,  $\tilde{h}_\mu^\nu=  {h}^{(1)\,\nu}_{\;\;\mu}$.  Here
${}^{(1)}\!\tilde{\mathcal{A}}_{\mu\nu}=-4\partial^{\rho}\partial^{\sigma}\tilde{C}_{\mu\rho\nu\sigma}$ 
is the first order form of $\mathcal{A}_{\mu\nu}$, defined in Eq.~\eqref{E:Amunu} for perturbed Minkowski spacetime. 
In our case, it has the explicit form
\begin{equation}
{}^{(1)}\!\tilde{A}_{\mu}{}^{\nu}(x)=\widetilde{\square}\widetilde{\square}\bigl[g(\eta)\hat{h}^{(1)\,\nu}_{\;\;\mu}(x)\bigr]\,,
\label{E:A-tilde}
\end{equation}
where $\widetilde{\square} = -{\partial}_{\eta}^2 +\nabla^2$ is the wave operator for flat spacetime.
 
  The action of the nonlocal kernel $H_{\lambda}(x-x')$ on a scalar function $f(x)$ is described by,
\begin{equation}\label{E:H-def}
	\int\!d^{4}x'\;H_{\lambda}(x-x')f(x')=
	\int\!d\Omega\int_{-\infty}^{0}\!du\;\left[\ln\left(-\frac{u}{\lambda}\right)\,\frac{\partial}{\partial u}+
	\frac{1}{2}\frac{\partial}{\partial v}\right]f(x')\bigg|_{v=0}\,,
\end{equation}
where $u$ and $v$ are null coordinates in $x'$ for radial null rays with origin at $x$. The integration $\displaystyle{\int d\Omega}$ 
is performed over the solid angle spanned by the past lightcone of the point $x$. The parameter $\lambda$ in the kernel 
$H_{\lambda}(x-x')$ arises from the ambiguity in a renormalized stress tensor; a shift of the value of $\lambda$ changes the 
constant $c_2$ in Eq.~\eqref{E:local}. Note that $\mathcal{P}_{\mu\nu}(x)$ is a local quantity, but $\mathcal{Q}_{\mu\nu}(x)$ is
nonlocal, and depends upon an integral over the past lightcone of the point $x$.
Now we may write the $\mathcal{O}(\kappa^3)$ part of the graviton field as
 \begin{equation}\label{E:h3a}
	\hat{h}^{(3)}_{\mu\nu}=2\kappa^{2}\!\int\!d^{4}x'\sqrt{-\gamma(x')}\;G_{R}(x,x')\,\Bigl\{\mathcal{P}_{\mu\nu}[\hat{h}^{(1)}]+
	\mathcal{Q}_{\mu\nu}[\hat{h}^{(1)}]\Bigr\}\,,
\end{equation}
and will treat the two contributions in succession

\subsection{Contribution Associated with $\mathcal{P}_{i}{}^{j}$}

In the  transverse, trace-free gauge, the tensor $\mathcal{P}_{\mu\nu}(x)$ defined in Eq.~\eqref{E:Pmunu} has only
spatial components. In de Sitter spacetime, it may be expressed as
\begin{equation}
	\hat{\mathcal{P}}_{i}{}^{j}(x)=\frac{4\pi\alpha}{a^{4}(\eta)}\left\{\frac{1}{\eta^{2}}\Bigl[\widetilde{\square}\hat{h}^{(1)\,j}_{\;\;i}(x)+
	2\,\hat{h}^{(1)\,j}_{\;\;i,\,\eta\eta}(x)\Bigr]+\frac{2}{\eta}\,{\partial}_{\eta}\widetilde{\square}\hat{h}^{(1)\,j}_{\;\;i}(x)-
	\ln(-H\eta)\widetilde{\square}\widetilde{\square}\hat{h}^{(1)\,j}_{\;\;i}(x)\right\}\,,
\end{equation}
 where $\hat{h}^{(1)\,j}_{\;\;i}(x)$   is the free graviton field on the de Sitter background given in Eq.~\eqref{E:h1}, 
 but now multiplied by a factor of the switching function $g(\eta)$. This leads to the mode expansion 
\begin{align}
	\hat{\mathcal{P}}_{i}{}^{j}(x)&=2\kappa\int\!\frac{d^{3}\mathbf{k}}{(2\pi)^{\frac{3}{2}}}\sum_{\lambda}
	\varepsilon_{i}{}^{j}(\mathbf{k},\lambda)\,\hat{a}_{\mathbf{k}\lambda}\,e^{i\,\mathbf{k}\cdot\mathbf{x}}\,
	v_{1k}(\eta)+\mathrm{H.c.}\,,
\label{E:P1-expand}	
\end{align}
where the function $v_{1k}(\eta)$ is given by
\begin{align}\label{E:v1}
	v_{1k}(\eta)&=\frac{4\pi\alpha}{a^{4}(\eta)}\biggl[\frac{1}{\eta^{2}}\Bigl(\frac{d^{2}}{d\eta^{2}}-
	k^{2}\Bigr)-\frac{2}{\eta}\Bigl(\frac{d^{3}}{d\eta^{3}}+k^{2}\frac{d}{d\eta}\Bigr)\biggr.\notag\\
	&\qquad\qquad\qquad\qquad\qquad\qquad\quad-\biggl.\ln(-H\eta)\Bigl(\frac{d^{2}}{d\eta^{2}}+k
	^{2}\Bigr)^{2}\biggr]\Bigl[g(\eta)u_k(\eta)\Bigr]\,.
\end{align}

The contribution of  $\mathcal{P}_{i}{}^{j}$ to the power spectrum comes from the contribution of $v_{1k}(\eta)$
to  $z_k(\eta_r)$ through the integral in Eq.~\eqref{E:zk}. Denote this contribution by  $z_{1k}(\eta_r)$. We are
primarily interested in contributions to $P_{13}$, which grow at least as rapidly with increasing duration of 
inflation, $|\eta_0|$, as the linear growth found for  $P_{22}$. Terms in the integrand of Eq.~\eqref{E:zk} which
oscillate as $\eta \rightarrow -\infty$ cannot produce such growth, and can be ignored here. Note that $u_k(\eta)$,
and hence $v_{1k}(\eta)$ are both proportional to ${\rm e}^{-i k\eta}$. Thus only the part of $\widetilde{G}_{R}(\eta_r,\eta;{k})$
which is proportional to ${\rm e}^{i k\eta}$ can produce a growing contribution to $z_{1k}(\eta_r)$. Write
\begin{equation}
	\widetilde{G}_{R}(\eta_r,\eta;{k})=
	\frac{H^{2}}{2(2\pi k)^{3}}\,\bigl(1+i\,k\eta_r\bigr)\bigl(i+k\eta\bigr) {\rm e}^{i\,k(\eta-\eta_r)}+
	\text{terms proportional to ${\rm e}^{-i\,k(\eta-\eta_r)}$}\,,
\end{equation}
and drop the second term in this expression. In addition, we note that $|k \eta| \gg 1$ in the region which gives the dominant 
contribution to the $\eta$-integration, and write
\begin{equation}\label{E:Ggrow}
	\widetilde{G}_{R}(\eta_r,\eta;{k}) \approx
	\frac{H^{2}}{16 \pi^3 k^2}\,\bigl(1+i\,k\eta_r\bigr)\, \eta\, {\rm e}^{i\,k(\eta-\eta_r)} \,.
\end{equation}
We may combine Eqs.~\eqref{E:P13}, \eqref{E:v1} and \eqref{E:Ggrow} to
write the dominant contribution of  $\mathcal{P}_{i}{}^{j}$ to the power spectrum as
\begin{equation}
\frac{\kappa^2}{2 \pi^2}\; {\rm Re} \left[ z_{1k}(\eta_r)\, u^*_k(\eta_r) \right] \propto
\int^{\eta_r}_{-\infty}  \frac{d\eta}{\eta} \left[\left(g +5\eta g' +3\eta^2 g''\right)
+2\ln(-H\eta)\, \left( 2\eta g' +3\eta^2 g'' +\eta^3 g''' \right)\right] \,. 
\end{equation} 

Now we need to examine the rate of growth for large $|\eta_0|$ of each of the contributions to the above integral.
Recall that $g(\eta_r)=1$ and $g(\eta) \approx 1$ for $\eta \agt \eta_0 = -|\eta_0|$, but $g$ and its derivatives vanish
faster than any inverse power of $\eta$ as $\eta \rightarrow -\infty$. Similarly, the  derivatives of  $g$ vanish at 
$\eta = \eta_r$.  Thus 
\begin{equation}
\int^{\eta_r}_{-\infty}  \frac{d\eta}{\eta}\, g(\eta) \sim \ln|\eta_0|
\end{equation}\,,
\begin{equation}
\int^{\eta_r}_{-\infty} d\eta\, g'(\eta) = g(\eta_r) =1\,,
\end{equation}
and
\begin{equation}
\int^{\eta_r}_{-\infty} d\eta\, \eta\, g''(\eta)  = \left[\eta\, g'(\eta)\right]^{\eta_r}_{-\infty} -  \int^{\eta_r}_{-\infty} d\eta\, g'(\eta) =-1\,.
\end{equation}
Similarly, we find
 \begin{equation}
\int^{\eta_r}_{-\infty} d\eta\, \ln(-H\eta)\,g'(\eta) = \left[\ln(-H\eta)\,g(\eta)\right]^{\eta_r}_{-\infty} - 
\int^{\eta_r}_{-\infty} d\eta\, \frac{g(\eta)}{\eta} \sim -\ln|\eta_0|\,,
\end{equation}
\begin{equation}
\int^{\eta_r}_{-\infty} d\eta\, \ln(-H\eta)\,\eta \, g''(\eta)   =  \left[\eta\,  \ln(-H\eta)\,g'(\eta)\right]^{\eta_r}_{-\infty} - 
\int^{\eta_r}_{-\infty} d\eta\,[\ln(-H\eta) +1]\, g'(\eta) \sim -\ln|\eta_0|\,,
\end{equation}
and
\begin{eqnarray}
\int^{\eta_r}_{-\infty} d\eta\, \ln(-H\eta)\,\eta^2 \, g'''(\eta)   &=&  \left[\eta^2\,  \ln(-H\eta)\,g''(\eta)\right]^{\eta_r}_{-\infty} - 
\int^{\eta_r}_{-\infty} d\eta\,[2 \eta\, \ln(-H\eta) +\eta]\, g'(\eta)   \nonumber \\
&\sim& -2\, \ln|\eta_0|\,.
\end{eqnarray}
Thus the contribution of  $\mathcal{P}_{i}{}^{j}$ to $P_{13}$ can only grow logarithmically with increasing $|\eta_0|$,
and is hence subdominant compared to $P_{22}$, which grows linearly.

\subsection{Contribution Associated with $\mathcal{Q}_{i}{}^{j}$}
\label{S:Qij}

Now we turn to the contribution to $P_{13}$ from the nonlocal tensor, $\mathcal{Q}_{i}{}^{j}$. This contribution
involves integrations with two retarded Green's functions, $G_{R}(x,x')$ and the function  $H_{\lambda}(x-x')$
defined in Eq.~\eqref{E:H-def}, The null coordinates used in this expression may be taken to be $u=\eta'-\eta -r$
and $v=\eta'-\eta +r$, where $r= \mathbf{x'}$, so that
\begin{equation}
\frac{\partial}{\partial u} = \frac{1}{2}\left(\frac{\partial}{\partial\eta'}-\frac{\partial}{\partial r}\right) \,, \qquad
\frac{\partial}{\partial v} =\frac{1}{2}\left(\frac{\partial}{\partial\eta'}+\frac{\partial}{\partial r}\right)\,.
\end{equation}
The function $f(x')$ in Eq.~\eqref{E:H-def} can be taken to be a graviton mode function of the form 
$f(x') = {\rm e}^{i\,\mathbf{k}\cdot\mathbf{x'}}\; F(\eta')$. We may choose the coordinates $\mathbf{x'}$ such
that the origin is at $\mathbf{x}=0$, so $r = |\mathbf{x'}|$, and  that the $z'$ axis is in the direction of
the wavevector $\mathbf{k}$. In this case, $\mathbf{k}\cdot\mathbf{x'} = k\, z' = k\,r \, c$, where $c = \cos \theta'$
is the cosine of the polar angle in these coordinates. Now we have $f(x') = {\rm e}^{i\, k \,r \,c}\; F(\eta')$, which 
leads to
\begin{equation}
\left(\frac{\partial f}{\partial u}\right)_{v=0} = \frac{1}{2}\left[ F'(\eta-r) - i\, k \,r \,c\, F(\eta-r)\right]\, {\rm e}^{i\, k \,r \,c} \,,
\end{equation}
and 
\begin{equation}
\left(\frac{\partial f}{\partial v}\right)_{v=0} = \frac{1}{2}\left[ F'(\eta-r) + i\, k \,r \,c\, F(\eta-r)\right]\,  {\rm e}^{i\, k \,r \,c} \,.
\end{equation}
Note that along the $v=0$ line, $u=-2 r$, so that $\int_{-\infty}^{0}\!du = 2 \int^{\infty}_{0}\!dr$. 
We combine these results and perform the angular integrations to write Eq.~\eqref{E:H-def} as
\begin{equation}\label{E:H-F}
	\int\!d^{4}x'\;H_{\lambda}(x-x')f(x')= \frac{2\pi}{k}\,  \int^{\infty}_{0}\!dr \left[ F(\eta-r)\; y_1(r) +  F'(\eta-r)\; y_2(r)\right]\,,
\end{equation}
where
\begin{equation}
y_1(r) = \frac{k r \cos kr - \sin kr}{r^2}\; \left[1 -2 \ln\left(\frac{2 r}{\lambda} \right)\right]\,,
\end{equation}
and
\begin{equation}
y_2(r) = \frac{\sin kr}{r}\; \left[1 +2 \ln\left(\frac{2 r}{\lambda} \right)\right]\,.
\end{equation}

We can now write $\mathcal{Q}_{i}{}^{j}$ in a mode expansion of the form of Eq.~\eqref{E:P1-expand}, but with 
$v_{1k}(\eta)$ replaced with $v_{2k}(\eta)$, given by
\begin{equation}
v_{2k}(\eta) = \frac{2\pi \alpha}{k\, a^4(\eta)}\; \int^{\infty}_{0}\!dr \left[ F(\eta-r)\; y_1(r) +  F'(\eta-r)\; y_2(r)\right]\,,
\label{E:v2k}
\end{equation}
where
\begin{equation}
F(\eta) =\left(\frac{d^{2}}{d\eta^{2}}+k^{2}\right)^{2}\left[g(\eta)u_k(\eta)\right]\,.
\label{E:F}
\end{equation}
If we insert Eq.~\eqref{E:v2k} for  $v_{k}(\eta)$   in Eq.~\eqref{E:zk}, the result is $z_{2k}(\eta_r)$, the nonlocal
contribution to  $z_{k}(\eta_r)$. We can use the fact that $F'(\eta-r) = -\partial/{\partial r}\; F(\eta-r)$ and perform
an integration by parts on the second term in Eq.~\eqref{E:v2k}, but only if we initially restrict the range of integration
to $r \geq \epsilon >0$. In this case, we have
\begin{equation}
\int^{\infty}_{\epsilon}\!dr \,F'(\eta-r)\, y_2(r) =  F(\eta-\epsilon)\, y_2(\epsilon)  + \int^{\infty}_{\epsilon}\!dr \,F(\eta-r)\, y'_2(r) \,.
\end{equation}
Note that $F(\eta-r)\, y_2(r)$ vanishes as $r \rightarrow \infty$, due in part to the presence of derivatives of $g(\eta-r)$ in
$F(\eta-r)$. If we use
\begin{equation}
y_1(r) + y'_2(r) = 2k\, \frac{\cos kr}{r}\,,
\end{equation}
then we have
\begin{equation}
v_{2k}(\eta) = \frac{2\pi \alpha}{k\, a^4(\eta)}\; \lim_{\epsilon \to 0}\; \left[F(\eta-\epsilon)\, y_2(\epsilon)  + 
2 k \int^{\infty}_{\epsilon}\!dr \,F(\eta-r)\; \frac{\cos kr}{r} \right]\,.
\label{E:v2}
\end{equation}
This limit is finite because the $\ln \epsilon $ terms from the lower limit of the $r$-integration and from $y_2(\epsilon)$
cancel one another.

Now we may use Eqs.~\eqref{E:grav-mode} and \eqref{E:F} to write $F(\eta)$ in terms of derivatives of the switching
function, $g(\eta)$, as
\begin{equation}
F(\eta) = c_0\, \left[-8ik^3\, g'(\eta) +4k^2\, (2-ik \eta)\, g''(\eta) +
4 k^2 \eta\, g^{(3)}(\eta) +(1+ik\eta)\,g^{(4)}(\eta)\right]\; {\rm e}^{-i\,k \eta} \,,
\label{E:F-2}
\end{equation}
where 
\begin{equation}
c_0 = i\,\frac{H}{(2k^{3})^{\frac{1}{2}}}\,.
\label{E:c0}
\end{equation}
Note that there are no terms in the above expression for $F(\eta)$ which are proportional to $g(\eta)$ itself. Recall that
$g(\eta)$ varies from $0$ to $1$ over an interval in $\eta$ of order $\Delta$, which may be either of the order of or less than
$|\eta_0|$. In any case, we require slow switching in the sense that $\Delta \gg 1/k$. We can estimate the magnitude of
the $n$-th derivative of
$g(\eta)$ as being of order $1/\Delta^n$. This means that the $g^{(3)}(\eta)$ and $g^{(4)}(\eta)$ terms in Eq.~\eqref{E:F-2}
are suppressed by factors of $k/\Delta$ and $(k/\Delta)^2$, respectively, compared to the $g''(\eta)$ term, and may be
ignored. We cannot assess the relative magnitudes of the  $g'(\eta)$ and  $g''(\eta)$ terms because of the  $k\eta$ term in the latter,
but we can assume that $|k\eta| \gg 1$, and shift the argument of $F$ to write
\begin{equation}
F(\eta-r) \approx c_0\, \left[-8ik^3\, g'(\eta-r) - 4ik^3\, (\eta-r)\, g''(\eta-r) + \cdots \right]\; {\rm e}^{-i\,k (\eta-r)} \,.
\label{E:Fa}
\end{equation}

As in the previous subsection, we are concerned with contributions to the power spectrum which grow as $|\eta_0|$ increases,
which can only come from non-oscillatory terms in the $\eta$ integration.   Because $F(\eta-r) \propto {\rm e}^{-i\,k (\eta-r)}$,
the form of $\widetilde{G}_{R}(\eta_r,\eta;{k})$ given in Eq.~\eqref{E:Ggrow} will contribute the dominant contribution.
Now we may combine Eqs.~\eqref{E:zk}, \eqref{E:Ggrow},  \eqref{E:v2},  \eqref{E:F-2}, and \eqref{E:Fa}  to write
\begin{eqnarray}
z_{2k}(\eta_r) &\approx& -16 \pi \,i \alpha \kappa^2 H^2\, c_0\, \bigl(1+i\,k\eta_r\bigr)\,  {\rm e}^{-i\,k \eta_r}\,
\int_{-\infty}^{\eta_{r}}\!d\eta \;g^{2}(\eta)\, \eta\;  \lim_{\epsilon \to 0}\; \biggl\{ [ 2g'(\eta) +\eta\, g''(\eta) ]\, y_2(\epsilon)   \nonumber \\
 &+& 2 k \int^{\infty}_{\epsilon}\!dr \, \frac{\cos kr}{r} \, {\rm e}^{i\,k r} \; \left[ 2g'(\eta-r) +(\eta -r)\, g''(\eta-r)     \right] \biggr\} \,.
\end{eqnarray}
The contribution of $z_{2k}(\eta_r)$ to the power spectrum is from the quantity
\begin{eqnarray}
&{\rm Re}& \left[ z_{1k}(\eta_r)\, u^\ast_k(\eta_r) \right]  = 32 \pi  \alpha \kappa^2 H^2\, |c_0|^2\, k\,\left(1+ \,k^2 \eta_r^2 \right) \times\nonumber \\
&&\int_{-\infty}^{\eta_{r}}\!d\eta \;g^{2}(\eta)\, \eta\; \int^{\infty}_{0}\!dr \, \frac{\cos kr}{r} \, \sin kr \; \left[ 2g'(\eta-r) +(\eta -r)\, g''(\eta-r)  \right] \,.
\label{E:zu}
\end{eqnarray}
Note that the $y_2(\epsilon)$ term and the $\ln \epsilon $ term from the lower limit of the $r$-integration in $z_{1k}(\eta_r)\, u^\ast_k(\eta_r)$
are pure imaginary, and do not contribute here. This allows us to extend the lower limit of the $r$-integration to zero. Next note that
\begin{equation}
\int^{\infty}_{0}\!dr \, \frac{\cos kr}{r} \, \sin kr = \frac{1}{2}\, \int^{\infty}_{0}\!dr \, \frac{\sin 2kr}{r} =\frac{\pi}{4}\,,
\end{equation}
and that the dominant contribution to this integral comes from $r\alt k^{-1}$. Because the dominant contribution to the $\eta$-integral
comes from $|\eta| \gg k^{-1}$, we may use $\eta -r \approx \eta$ in Eq.~\eqref{E:zu}. We may combine this result with Eqs.~\eqref{E:P13}
and \eqref{E:c0} to write the contribution of the nonlocal tensor, $\mathcal{Q}_{i}{}^{j}$ to the power spectrum as
\begin{equation}
P_{13}(k) = \frac{\alpha \, \kappa^4\, H^4}{\pi \, k^2} \, \left(1+ \,k^2 \eta_r^2 \right) \,
 \int_{-\infty}^{\eta_{r}}\!d\eta \;g^{2}(\eta)\, \eta\, \left[ 2g'(\eta) + \eta \, g''(\eta)  \right] \,.
 \label{E:P13f}
\end{equation}
The integral in the above expression will be shown below to grow at least linearly as $|\eta_0|$ increases. Hence it is the dominant
contribution to $P_{13}$, as the $\mathcal{P}_{i}{}^{j}$ contribution grows only logarithmically.

It is of interest to compare the structures of the expressions for $P_{22}$, Eq.~\eqref{E:P22}, and for $P_{13}$, Eq.~\eqref{E:P13f}.
We note that $P_{22}$ contains four powers of the switching function $g(\eta)$. This arises because $P_{22}$ is a double integral of the
stress tensor correlations function, Eq.~\eqref{E:P22-0}, and each stress tensor operator contributes a factor of $g^2$. In contrast,
$P_{13}$ contains three powers of $g$. Of these, a factor of  $g^2$ arises in Eq.~\eqref{E:h3-1}, the relation between 
$ \langle{}^{(1)}\hat{T}_{\mu\nu}[\hat{h}^{(1)}] \rangle_M$ and $\langle \hat{h}^{(3)}_{\mu\nu} \rangle_M$. The remaining factor involving
derivatives of $g(\eta)$ comes from the operator $\hat{h}^{(1)}$ upon which $\langle {}^{(1)}\hat{T}_{\mu\nu}[\hat{h}^{(1)}] \rangle_M$ depends.
There is nominally a fourth factor of $g$ in $P_{13}$, which like $P_{22}$, is proportional to $\kappa^4$, but this is a factor of
$g(\eta_r)=1$ coming from $\hat{h}^{(1)}(\eta_r)$. Thus we see that for general switching functions, it is not possible for $P_{22}$
and $P_{13}$ to cancel one another.

\begin{figure}[htbp]
	\centering
		\includegraphics[scale=0.8]{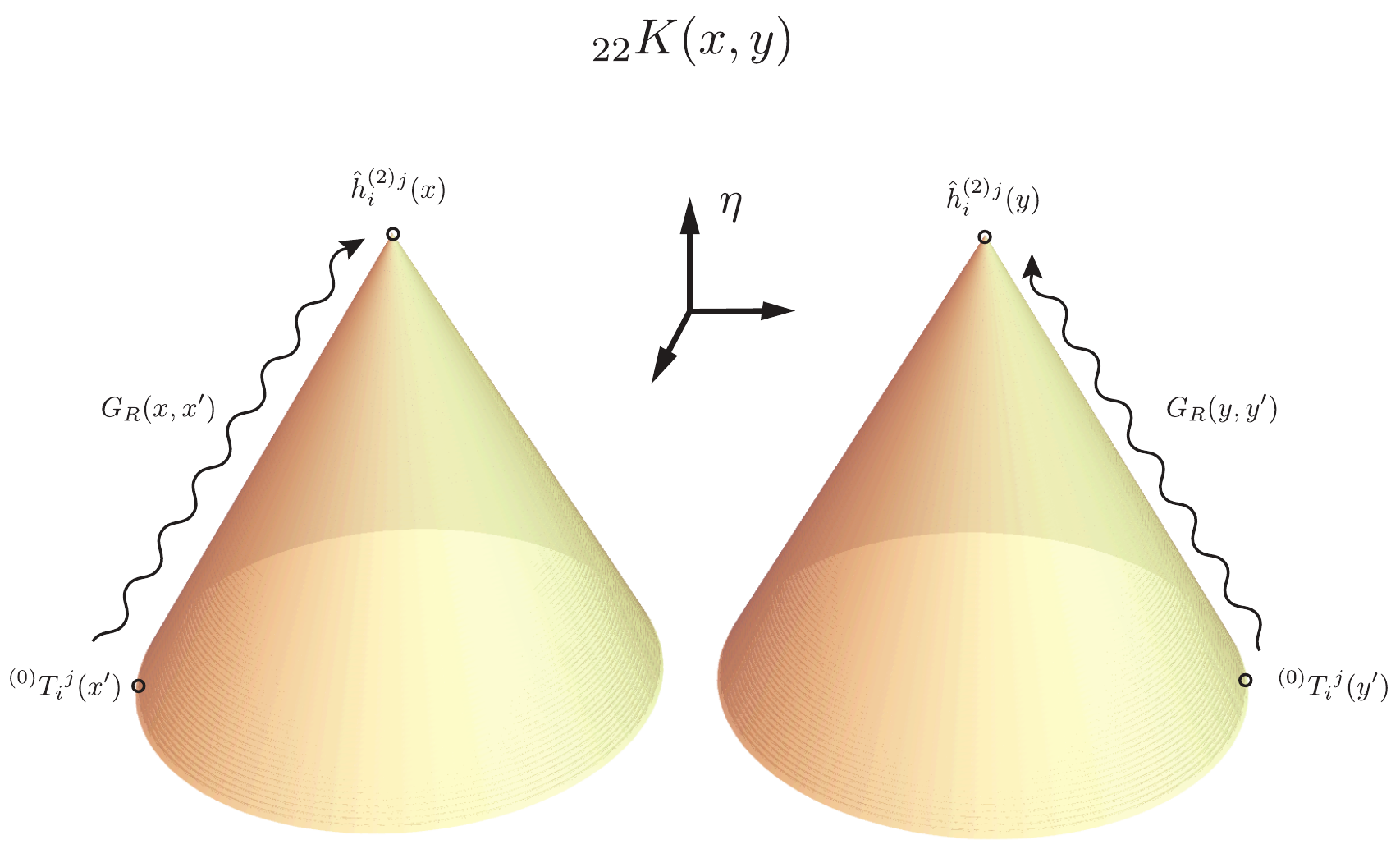}
		\caption{ The structure of the correlation function ${}_{22}K(x,y)$, which leads to  $P_{22}$ is illustrated. It involves an integral over the 
		past lightcones
		of points $x$ and $y$ of the product of a retarded Green's function and the fluctuating part of the free electromagnetic field stress
		tensor, ${}^{(0)}\hat{T}_{i}{}^{j}$.}
	\label{fig:P22}
\end{figure}

\begin{figure}[htbp]
	\centering
		\includegraphics[scale=0.8]{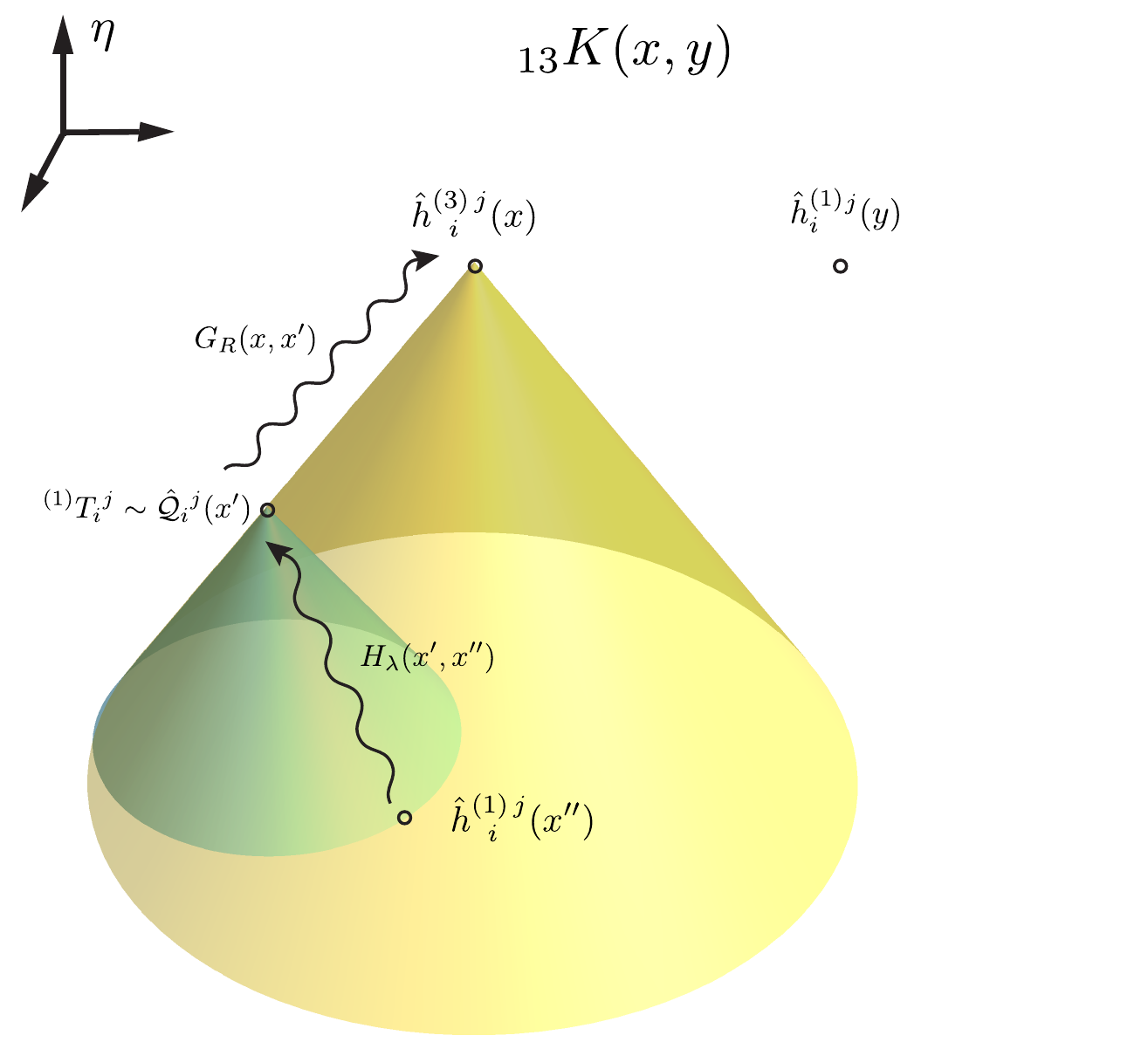}
		\caption{The structure of the correlation function ${}_{13}K(x,y)$, which leads to  $P_{13}$ is illustrated. It is the correlation function
		of the free graviton field, $\hat{h}^{(1)\;j}_{\;\;\; i}(y)$ with $\hat{h}^{(3)\;j}_{\;\;\; i}(x)$, which is given by an double integral over the 
		interior of the past lightcone of $x$.   }
	\label{fig:P13}
\end{figure}

It is also of interest to compare the domains of integration in the spacetime integrations which contribute to  $P_{22}$ and to $P_{13}$.
Recall that $P_{22}$ is the spatial Fourier transform of ${}_{22}K(x,y)$,  the equal time correlation function of $\hat{h}^{(2)\;j}_{\;\;\; i}(x)$ 
with itself, and that  $\hat{h}^{(2)\;j}_{\;\;\; i}(x)$ is given by Eq.~\eqref{E:h2soln}, which is an integral over the past lightcone of the spacetime point
$x$. This situation is illustrated in Fig.~\ref{fig:P22}. In contrast,  $P_{13}$ is the spatial Fourier transform of  ${}_{13}K(x,y)$, the equal time 
correlation function of the free graviton field, $\hat{h}^{(1)\;j}_{\;\;\; i}(y)$, and the third order correction to the graviton field, 
$\hat{h}^{(3)\;j}_{\;\;\; i}(x)$. The latter is given by Eq.~\eqref{E:h3-1}, an integral over the past lightcone of $x$ of the first order
correction to the stress tensor, ${}^{(1)}\hat{T}_{\mu\nu}[\hat{h}^{(1)}](x')$. The dominant contribution to the latter is the nonlocal
tensor $\mathcal{Q}_{i}{}^{j}$, which is in turn given by Eq.~\eqref{E:Qmunu}, an integral over the past lightcone of $x'$ of a linear
functional of the free graviton field,  $\hat{h}^{(1)\;j}_{\;\;\; i}(x'')$. The net result is an integration over the interior of the past lightcone 
of $x$. This is illustrated in Fig.~\ref{fig:P13}. This also serves to illustrate that the structures of  $P_{22}$ and to $P_{13}$ are quite
different.

\section{Combined Power Spectra}
\label{S:combined}

In this section, we discuss the combined $O(\kappa^4)$ contribution to the power spectrum, $P_4 (k)= P_{22} + P_{13}$, and its correction 
to the free graviton power spectrum $P_{11}(k)$. We restrict attention to the case where the conformal matter field is the 
electromagnetic field, for which
 \begin{equation}
\alpha = \frac{1}{320 \pi^3} \,.
\end{equation}
We may extend the upper limit of the integration in Eq.~\eqref{E:P13f} to $\eta =0$, as the range $\eta_r <\eta <0$ will give a subdominant
contribution for large $|\eta_0|$. After doing this, we may perform an integration by parts and use Eq.~\eqref{E:kappa} to write $P_{13}$ as
 \begin{equation}
P_{13}(k) = -\frac{2\,\ell_p^4\, H^4}{5 \pi^2 k^2} \, \left(1+ \,k^2 \eta_r^2 \right)\,
 \int_{-\infty}^{0}\!d\eta \;g^{2}(\eta)\, \eta^2\, [g'(\eta)]^2  \,.
  \label{P13f2}
\end{equation}
Note that $P_{13}(k) < 0$.
 We may similarly extend the upper limit of the integral in the expression for $P_{22}$, Eq.~\eqref{E:P22}, and write the combined
 power spectrum as
 \begin{equation}
P_4(k) =  P_{22} + P_{13} = \frac{\,\ell_p^4\, H^4}{10 \pi^2 k^2} \, \left(1+ \,k^2 \eta_r^2 \right)\; {\cal I}  \,,
\end{equation}
 where 
\begin{equation}
{\cal I}  =  \int_{-\infty}^{0}\!d\eta \;g(\eta)\, \left\{ g^3(\eta) -4 \eta^2\, [g'(\eta)]^2 \right\} \,.
\end{equation} 
The first term in the above integrand comes from the positive $P_{22}$ part, and the second term form the negative $P_{13}$ part.

 \subsection{ Specific Choices of the Sampling Function}
 \label{sec:choices}
 
 Next we examine results for several explicit choices for the switching function $g(\eta)$.
 
 \subsubsection{Exponential Switching}
 
 Let
 \begin{equation}
g(\eta) = {\rm e}^{p\, \eta}\,,
\label{E:exp-switch}
\end{equation}
where $p>0$. In this case, the effective Newton's constant is exponentially damped as $\eta \rightarrow - \infty$, and we
find
\begin{equation}
{\cal I}  =-\frac{5}{108p} \approx -0.046\, p^{-1}\,.
\label{E:I-exp}
\end{equation}
 Here $|\eta_0| = p^{-1}$ is the approximate duration of the switching in conformal time, so we have linear growth
 of $P_4$. In addition, $P_4 < 0$, corresponding to a reduction in the total graviton power spectrum.  
 
 A more general exponential-type switching function is
 \begin{equation}
g(\eta) = {\rm e}^{-|p\, \eta |^b}\,,
\end{equation}
 which leads to 
 \begin{equation}
{\cal I}  =- \frac{1}{54p}\left[8 b\times3^{\frac{b-1}{b}}\Gamma \left(\frac{2 b+1}{b}\right)
-27\times2^{\frac{b-2}{b}}\Gamma \left(\frac{b+1}{b}\right)\right]<0\,.
\end{equation}
A plot reveals that this function is negative for all values of $b$, and includes Eq.~\eqref{E:I-exp} for the case
$b =1$, and Gaussian switching when   $b =2$,

 \subsubsection{Lorentzian Switching}

A more gradual form of switching arises from a Lorentzian function,
\begin{equation}
g(\eta)=\frac{1}{1+(\eta/|\eta_0| )^{2}}\,,
\end{equation}
which yields
\begin{equation}
{\cal I}  =-\frac{\pi}{32}\,|\eta_0| \,.
\end{equation}
Again  $P_4 < 0$, and its magnitude grows linearly with increasing $|\eta_0|$.

 \subsubsection{A Polynomial, Finite Duration Switching Function}
 
 Here we wish to consider a function $g(\eta)$ which is strictly zero before a certain time, but for which both $g(\eta)$
 and its first three derivative are continuous. This insures that $g$ and its first four derivatives are finite everywhere.
 One such function is 
 \begin{equation}
			g(\eta)=\begin{cases}
						1\,,&\eta_{0}+\Delta\leq\eta\leq0\,,\\
						g_{P}(\tau)\,,&\eta_{0}\leq\tau\leq\eta_{0}+\Delta\,,\\
						0\,,&\eta<\eta_{0}\,,
						\label{E:poly}
					\end{cases}
		\end{equation}
where $g_{P}(\eta)$ is a polynomial given by
\begin{equation}
g_{P}(\eta)=-\frac{1}{\Delta^{7}}\left(\eta-\eta_{0}\right)^{4}\left[20\left(\eta-\eta_{0}\right)^{3}-70\Delta\left(\eta-\eta_{0}\right)^{2}+84\Delta^{2}\left(\eta-\eta_{0}\right)-35\Delta^{3}\right]\,.
\end{equation} 
 Here $0 <  \Delta < |\eta_0|$, so $\eta_0 < \eta_{0}+\Delta < 0$. The switch-on begins at $\eta = \eta_0$, and ends at
 $\eta =  \eta_{0}+\Delta$, so $\Delta$ is the duration of the switch-on. 
 
 The result here is
  \begin{equation}
{\cal I}  =-\frac{1400 \eta_{0} ^2}{429 \Delta }-\frac{19599 \eta_{0} }{4199}-\frac{483928444 \Delta }{277272567}<0\,.
\end{equation}
There are two limits of interest here. First we can hold the ratio $\Delta/|\eta_0|$ fixed as $|\eta_0|$ becomes large.
In this case, the magnitude of $P_4$ grows linearly in $|\eta_0|$, as in the previous examples. A second possibility
is to hold $\Delta$ fixed as $|\eta_0|$ becomes large, in which case the magnitude of $P_4$ grows quadratically in $|\eta_0|$.
The second option corresponds to a fixed switching interval in conformal time, followed by an increasing interval of inflation.

 \subsubsection{A $C^\infty$ Finite Duration Switching Function}

Now we examine a function qualitatively similar to the previous example, but which is infinitely differentiable. Let
\begin{equation}
g(\eta)=\begin{cases}
			{\rm e}^{-\frac{\Delta}{\eta_0}} \, {\rm e}^{-\frac{\Delta}{\eta-\eta_{0}}}\,,&\eta_{0}\leq\eta<0\,,\\
			0\,,&\eta\leq\eta_{0}\,.
			\label{E:compact}
		\end{cases}
\end{equation}
As in the previous example, this function switches on at $\eta = \eta_0$ and reaches $g=1$ at $\eta = 0$. If $\Delta < |\eta_0|$,
the approximate duration of the switch-on is about $\Delta$. In this case, we find
\begin{equation}
{\cal I}  = -\frac{8\eta^{2}_{0}}{27\Delta} + |\eta_0| + 4\Delta\, {\rm Ei}\left(-4\,\frac{\Delta}{ |\eta_0|} \right) \,, 
\label{E:I-compact}
\end{equation}
where ${\rm Ei}$ denotes the exponential integral function.
This result has the same general behavior found in the previous subsection. If  $|\eta_0|$ becomes large for fixed $\Delta$, then
we again have quadratic growth:
\begin{equation}
{\cal I}   \sim -\frac{8\eta^{2}_{0}}{27\Delta} \,.
\end{equation}
If we let $\xi = \Delta/|\eta_0|$, then Eq.~\eqref{E:I-compact} may be written as
\begin{equation}
{\cal I}  = |\eta_0|\; F(\xi) \,,
\end{equation}
where 
\begin{equation}
F(\xi) = -\frac{8}{27 \, \xi} +1 + 4 \xi {\rm e}^{4 \xi} \; {\rm Ei}(-4 \xi) \,.
\end{equation}
This result holds for all $\xi$. The asymptotic forms of $F(\xi)$ are
\begin{equation}
F(\xi) \sim  -\frac{8}{27 \, \xi}\,, \qquad \xi \ll 1\,, 
\end{equation}
and
\begin{equation}
F(\xi) \sim  -\frac{5}{108 \, \xi}\,, \qquad \xi \gg 1\,.
\end{equation}
A plot  indicates that $F(\xi) < 0$  for all intermediate values. Thus if $\xi$ is fixed and $|\eta_0|$ becomes large, then $P_4 <0$
and its magnitude grows linearly with  $|\eta_0|$, as in the previous example.

 \subsection{ Switching in Comoving Time}

  It has been convenient in much of our analysis to use the conformal time $\eta$ as the time coordinate. However, the proper 
  time for comoving observers is $t$, the comoving time. During inflation, when $a(t) = H\, {\rm e}^{- H t}$, these time coordinates 
  are related by
 \begin{equation}
\eta = \int \frac{dt}{a(t)} = -  \frac{1}{H}\, {\rm e}^{- H t} \,.
\end{equation}
Let inflation end at $t=t_r=0$, or $\eta =\eta_r = - H^{-1}$. If the universe expands by a factor of ${\rm e}^{N}$, then inflation
begins at about $\eta =\eta_0 = -H^{-1}\, {\rm e}^{N}$, but at $t= t_0 = -N\, H^{-1}$, a very large range of conformal time
corresponds to a much smaller range of comoving time. In addition, much of the initial change in $\eta$ corresponds to a small
fraction of the change in $t$. Let $\eta_{1/2} = \eta_0/2$ be when roughly one-half of the total conformal time interval
has elapsed, and let $t_{1/2}$ be the corresponding comoving time. The elapsed comoving time interval is 
$(t_{1/2}-t_0) = \ln 2 \,H^{-1} \approx 0.69  \,H^{-1}$. Thus, regardless of the size of $N$, the first one-half of the total conformal 
time corresponds to less than one e-fold time in comoving time.

\begin{figure}[htbp]
	\centering
		\includegraphics[scale=0.4]{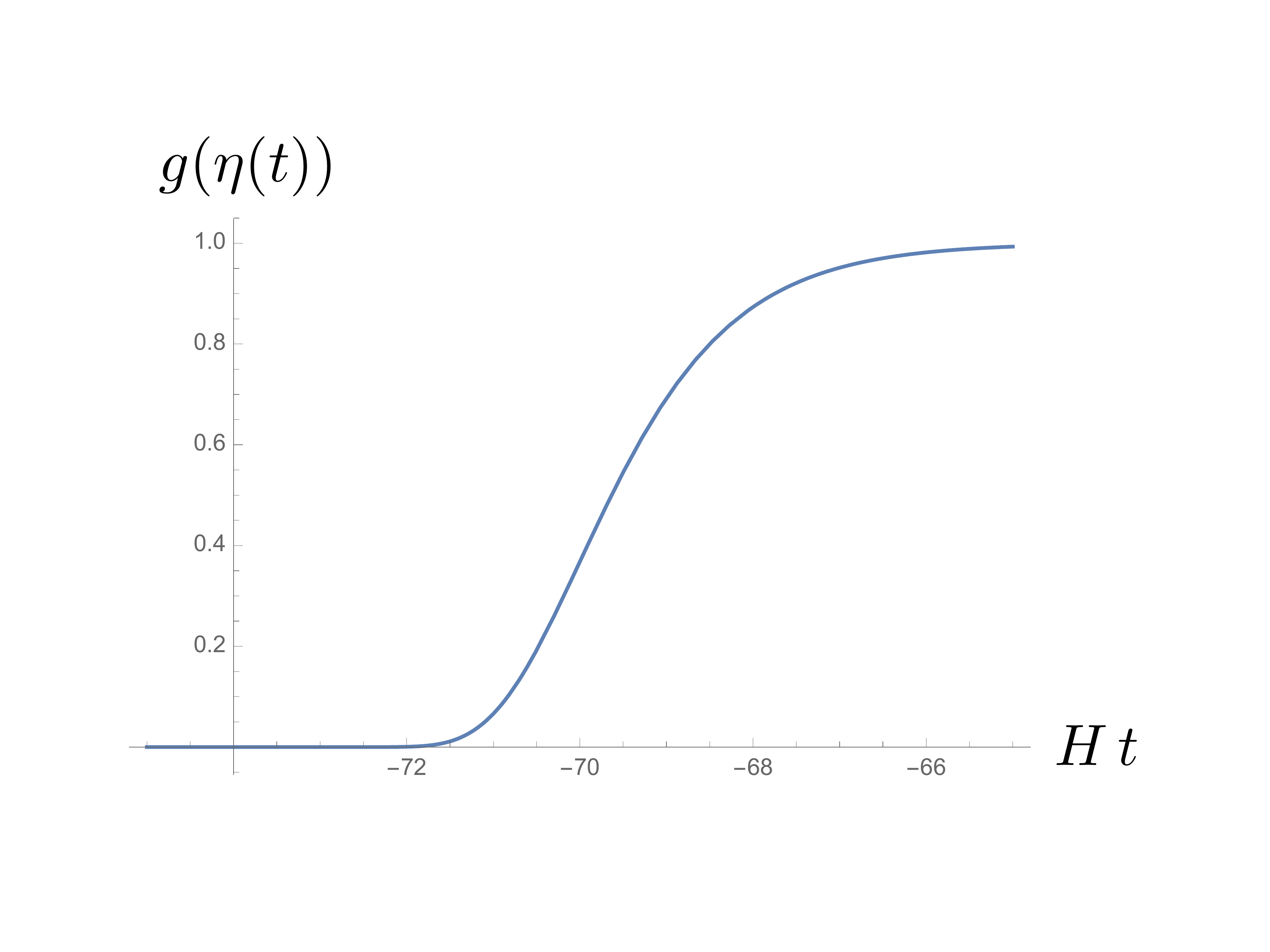}
		\caption{The exponential switching function, Eq.~\eqref{E:exp-switch-t} is plotted as a function of comoving time
		$t$ for the case  $p= H/70$. This graph illustrates that $g(\eta(t))$ rises from zero to near one in a comoving time 
		interval of  $\Delta t \approx 6\, H^{-1}$.  }
	\label{fig:t-switch}
\end{figure}

The implication is that apparently slow switching in conformal time is relatively rapid switching in comoving time.
We can illustrate this more explicitly for the case of the exponential switching function, Eq.~\eqref{E:exp-switch}, which we 
can express as a functions of $t$:
\begin{equation}
g(\eta(t)) = \exp[-p\, H^{-1}  {\rm e}^{- H t} ]\,.
\label{E:exp-switch-t}
\end{equation} 
The switch-on part of this function is plotted in Fig.~\ref{fig:t-switch} for the case $p= H/70$, corresponding to
$\eta_0 = 70 \, H^{-1}$ or $N=70$ We can see that essentially all of the switch-on occurs in a comoving time interval
of $\Delta t \approx 6\, H^{-1}$.

This is the comoving time required for about six e-folds. This seems to be a plausible switching interval. However, the
functions discussed in Sec.~\ref{sec:choices} which describe more rapid switching, such as Eqs.~\eqref{E:poly} 
or \eqref{E:compact} with $\Delta \ll |\eta_0|$, can correspond to a comoving switching time $\Delta t \ll H^{-1}$.
This seems unphysically short. Recall that these rapid switching cases are also those which give $P_4$ growing 
quadratically in $|\eta_0|$. Thus the quadratic growth is probably an artifact of a too rapid switch-on. This leaves
the linear growth behavior as the more reasonable case.

 \subsection{ Some Estimates}
 \label{S:estimates}
 
 Here we consider some numerical estimates for the correction to the graviton power spectrum. Assume that 
 ${\cal I} = -\beta\, |\eta_0|$, where $\beta$ is a numerical constant of order one, or somewhat less, determined by the switching
 function. The total power spectrum, including the quantum corrections computed in previous sections, becomes
 \begin{equation}
P_T(k) = P_{11} + P_4\,.
\end{equation}
It is slightly reduced from the free graviton spectrum $P_{11}(k)$ by the factor
\begin{equation}
R = \frac{P_T}{P_{11}} = 1 - \frac{\beta \, \ell_p^2\, H^2}{20} \, \left(\frac{k}{H} \right)\, S\,,
\label{E:R}
\end{equation}
where $S = H\, |\eta_0|$ is the total scale factor increase during inflation. Here we are setting the scale factor to
one at the end of inflation, so $a(\eta_r)=1$ and $\eta_r =- H^{-1}$.

During inflation, the Friedmann equation gives
 \begin{equation}
H^2 = \frac{8 \pi}{3}\, \ell_p^2\;  \rho_V\,,
\end{equation}
where $\rho_V$ is the vacuum energy density. Consider a model with efficient reheating at the end of inflation, so
the reheating temperature $T_R$ is given by $\rho_V = \pi^2\, T_R^4/15$, in units where Boltzmann's constant is one.
In this model, we have
 \begin{equation}
\ell_p\, H = 1.5 \times 10^{-4} \; \left( \frac{T_R}{10^{17}\, {\rm GeV}} \right)^2   \,.
\end{equation}

Consider a gravity wave whose proper wavelength today is $\lambda_0$, so its comoving wavenumber is
\begin{equation}
k = a_0 \, \frac{2 \pi}{\lambda_0}\,,
\end{equation}
where $a_0 \approx T_R/ 3{\rm K}$ is the present scale factor. Let $\lambda_0 = f\, d_H$, where $d_H \approx 
1.3 \times 10^{28} \,{\rm cm}$ is the current horizon size. We then find
\begin{equation}
\frac{k}{H} = \frac{2.1 \times 10^{-27}}{f} \, \left( \frac{10^{17}\, {\rm GeV}}{T_R} \right)\,.
\end{equation}
We are interested in perturbations for which $f < 1$, so $k^2 \, \eta_r^2 \ll 1$ for all realistic choices for $T_R$,
and the free graviton power spectrum, Eq.~\eqref{E:P11-2}, becomes
\begin{equation}
P_{11}(k) = \frac{2\, \ell_p^2\, H^2}{\pi^2 \, k^3}\,,
\end{equation}
or
\begin{equation}
{\cal P}_{11} = \frac{2}{\pi^2} \, \ell_p^2\, H^2 \approx 4.5 \times 10^{-9}\, \left(\frac{T_R}{10^{17}\, {\rm GeV}} \right)^4\,.
\end{equation}

Now we may write the magnitude of the fractional change in the power spectrum as
\begin{equation}
|R-1| = 2.4 \times 10^{-10}\, \beta\, \left(\frac{10^{-3}}{f}\right) \, \left(\frac{S}{10^{23}}\right) \, 
\left(\frac{T_R}{10^{17}\, {\rm GeV}} \right)^4 \,.
\end{equation}
This change will be very small unless $S$ is much larger than the minimal value of about $10^{23}$ needed to solve the 
horizon and flatness problems. However, if $S$ becomes too large, then the perturbation in question will have a proper
wavelength below the Planck length at the beginning of inflation. This initial wavelength can be expressed as
\begin{equation}
\lambda_i = \frac{\lambda_0}{S}\, \left(\frac{3 {\rm K}}{T_R} \right)\,.
\end{equation}
The status of transplanckian frequency modes is controversial. They play a crucial role in the Hawking's derivation
of black hole particle creation~\cite{Hawking}, but it seems questionable that perturbation theory holds for such modes. 
If we impose the requirement of no transplanckian modes, so $\lambda_i \geq \ell_p$, then we find
\begin{equation}
\frac{S}{f} \leq 2 \times 10^{31}\,  \left( \frac{10^{17}\, {\rm GeV}}{T_R} \right) \,,
\end{equation}
and
\begin{equation}
|R-1| \leq 4.8 \times10^{-5}\, \beta \, \left(\frac{T_R}{10^{17}\, {\rm GeV}} \right)^2\,.
\end{equation}
In this case, the decrease in the graviton power due to the effects calculated in this paper will be fairly small.

\section{Summary and Discussion}
\label{S:sum}

In this paper, we have examined $O(\ell_p^4)$ quantum corrections to the tensor power spectrum in inflationary models.
The corrections with which we are concerned come from the coupling of gravitons to a conformal matter field, which we take
to be the electromagnetic field. There are two distinct corrections to the power spectrum. One is $P_{22}$, which arises
from vacuum stress tensor fluctuations of the matter field. One can view this contribution as being the gravity waves
radiated by the fluctuating stress tensor, or equivalently, the passive fluctuations of the gravitational field driven by the
quantum stress tensor fluctuations. The other contribution is  $P_{13}$, which arises from a modification of the graviton
field in de Sitter spacetime, produced by its coupling to the renormalized expectation value of the matter field. This can 
be viewed as a correction to the active fluctuations of the quantized tensor perturbations of  de Sitter spacetime.

A key feature of our approach is the use of the switching function, $g(\eta)$.  This function may be viewed as describing
a switching of the coupling of gravity with the matter field through a time-dependent Newton's constant. It may
also be viewed as a form of time averaging of the quantum stress tensor operator, which is essential for a
meaningful treatment of quantum stress tensor fluctuations. The viewpoint adopted in this paper is that this averaging
is more than a formal device, and represents actual physical processes associated with the measurement of the
stress tensor of a quantized field. In the case of cosmology, we postulate that it describes some physical effects in
the early universe. These effects are presently not well understood and are associated with a choice of initial conditions.

We find that combined correction, $P_4 = P_{22} + P_{13}$, depends upon the choice of switching function. This seems to
be required in our approach, because $P_{22}$ and $P_{13}$ scale with different powers of $g(\eta)$, as discussed 
at the end of Sec.~\ref{S:P13}.  For
all of our explicit choices of  $g(\eta)$, we find $P_4 < 0$, so the effect is a  slight reduction in the  tensor power spectrum,
compared to the result obtained from consideration of free gravitons in de Sitter spacetime. We also find that the
magnitude of $P_4$ is proportional to the scale factor change during inflation. This is not due to growth as inflation
progresses, but rather is due to decreasing proper wavelength of the perturbation at the initial time, as this time is made 
earlier. Recall that we consider a perturbation with given proper wavelength today, so its  proper wavelength at the beginning
of inflation depends upon the amount of subsequent expansion. Furthermore, the dominant contributions to $P_{22}$ and 
$P_{13}$ come from the beginning of inflation, or the initial switch-on interval, as evidenced by their growth with
increasing $|\eta_0|$, the effective switch-on time in conformal time. The essential physical reason for this behavior is
that quantum stress tensor fluctuations are greater on smaller length scales. This introduces a breaking of de Sitter symmetry
by the initial conditions. The situation described here is quite different from the usual behavior in inflationary models, where
classical perturbations of a given proper wavelength at the beginning of inflation are more effectively redshifted away by
an increasing duration of inflation.

Our conclusions clearly differ from those of  Fr{\"o}b, {\it et al.}~\cite{FRV12}, who find no significant dependence of  $P_4$
upon the initial conditions. This seems to be due to physical inequivalence of our approaches. Fr{\"o}b, {\it et al.}~\cite{FRV12}
use a rather formal prescription, which they call an ``$i \epsilon$ prescription"  to select a de Sitter invariant state for the
coupled conformal field - gravity system. Our view is that this prescription is not physically well motivated. In light of the
discussion in the previous paragraph, we argue that one should expect to find the behavior found in this present paper, as well 
in Refs.~\cite{WKF07,FMNWW10,WHFN11}, where the quantum corrections to the power spectrum depend upon initial
conditions.

Our main result is a small decrease in the power spectrum of tensor perturbations which has the linear dependence upon
$k$ given in Eq.~\eqref{E:R}. Note that this has a distinct functional form from the usual spectral tilt due to weak dependence
of $H$ upon $k$. The latter effect is due to the effective value of $H$ varying as different length scales leave the horizon
during inflation, which will also be present in our model. This is usually modeled by a factor of the form $k^{n_T}$, where
$n_T$ is the tensor spectral index. The estimates given in Sec.~\ref{S:estimates} indicate that the quantum corrections to 
the power spectrum are small. However, if they can be observed, they could lead to insights about the initial conditions in
the early universe.

\section{acknowledgments}
We would like to thank Markus Fr{\"o}b, Albert Roura, Enric Verdaguer, and Richard Woodard for useful conversations.
This work was supported in part  by the U.S. National Science Foundation under Grant PHY-1607118, and by the 
Ministry of Science and Technology, Taiwan, ROC under Grant No. MOST104-2112-M-001-039-MY3.

\appendix
\section{Stress Tensor Correlation Function of the Electromagnetic Field in Minkowski Spacetime}
Given the stress tensor of the EM field
\begin{equation}\label{E:fnewr}
	T_{\mu\nu}=F_{\mu}{}^{\alpha}F_{\alpha\nu}-\frac{1}{4}\,g_{\mu\nu}\,F_{\alpha\beta}F^{\alpha\beta}\,,
\end{equation}
with $F_{\alpha\beta}=\partial_{\alpha}A_{\beta}-\partial_{\beta}A_{\alpha}$, we may define the correlation function between the stress tensor by 
\begin{equation}
	C_{\mu\nu\rho\sigma}(x,x')=\langle:T_{\mu\nu}(x):\,:T_{\rho\sigma}(x'):\rangle\,.
\end{equation}
It is convenient to use the Wick's expansion to express this correlation function in terms of that of the vector potential
\begin{equation}
	D_{\mu\nu}(x,x')=\langle A_{\mu}(x)A_{\nu}(x')\rangle=-\eta_{\mu\nu}D(x,x')\,,
\end{equation} 
if we choose the Lorentz gauge and the space is flat and unbounded, and $D(x,x')$ is the corresponding correlation function of the minimally 
coupled, massless scalar field. Here $\eta_{\mu\nu}$ is the Minkowski metric with $\eta_{\mu\nu}={\rm diag}(-1,+1,+1,+1)$.

When the EM field is in its vacuum state, the stress tensor correlation function takes the form~\cite{FW03}
\begin{align}
	C_{\mu\nu\rho\sigma}&=4\left(\partial_{\mu}\partial_{\nu}D\right)\left(\partial_{\rho}\partial_{\sigma}D\right)+
	2\eta_{\mu\nu}\left(\partial_{\rho}\partial_{\lambda}\right)\left(\partial_{\sigma}\partial^{\lambda}D^{(0)}\right)+
	2\eta_{\rho\sigma}\left(\partial_{\mu}\partial_{\lambda}\right)\left(\partial_{\nu}\partial^{\lambda}D\right)\notag\\
	&-2\eta_{\mu\sigma}\left(\partial_{\nu}\partial_{\lambda}D\right)\left(\partial_{\rho}\partial^{\lambda}D\right)-
	2\eta_{\nu\sigma}\left(\partial_{\mu}\partial_{\lambda}D\right)\left(\partial_{\rho}\partial^{\lambda}D\right)-
	2\eta_{\nu\rho}\left(\partial_{\mu}\partial_{\lambda}D^{(0)}\right)\left(\partial_{\sigma}\partial^{\lambda}D\right)\notag\\
	&-2\eta_{\mu\rho}\left(\partial_{\nu}\partial_{\lambda}D\right)\left(\partial_{\sigma}\partial^{\lambda}D\right)+
	\Bigl(\eta_{\mu\rho}\eta_{\nu\sigma}+\eta_{\mu\sigma}g_{\nu\rho}-
	\eta_{\mu\nu}\eta_{\rho\sigma}\Bigr)\left(\partial_{\lambda}\partial_{\kappa}D\right)\left(\partial^{\lambda}\partial^{\kappa}D\right)\,.
	\label{E:euinkds}
\end{align}
In particular, the $xyxy$--component of Eq.~\eqref{E:euinkds} is explicitly given by
\begin{align}
	C(x,x')=C_{xyxy}(x,x')&=\Bigl[\partial_{t}^{2}D(x,x')\Bigr]^{2}+\Bigl[\partial_{z}^{2}D(x,x')\Bigr]^{2}-
	\Bigl[\partial_{x}^{2}D(x,x')\Bigr]^{2}-\Bigl[\partial_{y}^{2}D(x,x')\Bigr]^{2}\notag\\
	&\qquad\qquad\qquad\qquad\qquad\qquad+2\Bigl[\partial_{x}\partial_{y}D(x,x')\Bigr]^{2}-2\Bigl[\partial_{t}\partial_{z}D(x,x')\Bigr]^{2}\notag\\
	&=\frac{2}{\pi^{4}}\frac{\bigl(\tau^{2}-Z^{2}\bigr)^{2}-X^{4}-Y^{4}+6X^{2}Y^{2}}{\bigl[\bigl(\tau-i\,\epsilon)^{2}-R^{2}\bigr]^{6}}\,,
\end{align}
where $D(x,x')$ is the Wightman function of the free massless scalar field $\phi$ in Minkowski space,
\begin{equation}
	D(x,x')=\langle\phi(x)\phi(x')\rangle=\frac{1}{4\pi^{2}}\frac{1}{-(\tau-i\,\epsilon)^{2}+R^{2}}\,,
	\label{E:Wightman}
\end{equation}
and we introduce the shorthand notations $R^{2}=X^{2}+Y^{2}+Z^{2}$ with $X=x-x'$, $Y=y-y'$,  $Z=z-z'$, and $\tau=t-t'$.

The spatial Fourier transformation of this two-point function is defined by
\begin{align}\label{E:dkfjejss}
	\widetilde{C}(\tau;\mathbf{k})&=\int\frac{d^{3}\mathbf{R}}{(2\pi)^{3}}\;C(\tau,\mathbf{R})\,e^{i\,\mathbf{k}\cdot\mathbf{R}}\,.
\end{align}
To simplify the calculations, we assume that $\mathbf{k}=k\,\hat{\mathbf{e}}_{z}$, and use the polar coordinate decomposition. Let 
$\rho^{2}=X^{2}+Y^{2}$, with $X=\rho\, \cos\phi$ and $Y=\rho\, \cos\phi$. We find the Fourier transformation Eq.~\eqref{E:dkfjejss} reduces to
\begin{align}
	\widetilde{C}(\tau;\mathbf{k})&=-\frac{1}{20\pi^{6}}\int_{-\infty}^{\infty}\!dZ\;e^{i\,k Z}\,
	\frac{1}{\bigl[\bigl(\tau-i\,\epsilon\bigr)^{2}-Z^{2}\bigr]^{3}}\,.\label{E:odfhbd}
\end{align}
Next we would like to rewrite the integrand in Eq.~\eqref{E:odfhbd} before evaluating the integral with the residue theorem,
\begin{align}
	\widetilde{C}(\tau;\mathbf{k})&=\frac{1}{40\pi^{6}}\int_{-\infty}^{\infty}\!dZ\;\left[\frac{d^{2}}{dZ^{2}}\frac{e^{i\,k Z}}
	{\bigl(Z-\tau+i\,\epsilon\bigr)^{3}}\right]\frac{1}{Z+\tau-i\,\epsilon}\,.
\end{align}
Since $k>0$, we can close the contour of $Z$ on the upper complex $Z$ plane, in which there is only one simple pole $Z=-\tau+i\,\epsilon$. 
Applying the residue theorem gives 
\begin{align}
	\widetilde{C}(\tau;\mathbf{k})&=\frac{1}{40\pi^{6}}\,2\pi\,i 
	\left[ \frac{d^{2}}{dZ^{2}}\frac{e^{i\,k Z}}{\bigl(Z-\tau+i\,\epsilon\bigr)^{3}}\right]_{Z=-\tau+i\,\epsilon}
	\notag\\
	&=-\frac{1}{1280\pi^{5}}\left(\frac{d^{2}}{d\tau^{2}}+k^{2}\right)^{2}\frac{i\,e^{-i\,k(\tau-i\,\epsilon)}}{\tau-i\,\epsilon}\,.\label{E:nveru}
\end{align}
The expression next to the differential operators can be recast into the form
\begin{align}
	\frac{i\,e^{-i\,\omega(\tau-i\,\epsilon)}}{\tau-i\,\epsilon}&=i\,e^{-i\,\omega(\tau-i\,\epsilon)}
	\left[\mathcal{P}\left(\frac{1}{\tau}\right)+
	i\,\pi\,\delta(\tau)\right]=\biggl[\frac{\sin\omega\tau}{\tau}-\pi\,\delta(\tau)\biggr]+i\,\mathcal{P}\left(\frac{\cos\omega\tau}{\tau}\right)\,.
\end{align}
Therefore the real part of Eq.~\eqref{E:nveru} will yield Eq.~\eqref{E:C}
\begin{align}
	\widetilde{C}(\tau;\mathbf{k})=\frac{1}{1280\pi^{5}}\left(\frac{d^{2}}{d\tau^{2}}+k^{2}\right)^{2}\left[-\frac{\sin k\tau}{\tau}+\pi\,\delta(\tau)\right]\,.
\end{align}
Note that this result differs in two ways from the correlation function given in Ref.~\cite{WHFN11}. First, we have corrected the overall numerical
coefficient. Second, we have included the delta function term, which arises from the use of $\tau -i \epsilon$, as opposed to $\tau$ , in the
Wightman functions, Eq.~\eqref{E:Wightman}.

\end{document}